\documentclass[twocolumn,10pt]{IEEEtran}

\usepackage{amsmath,amssymb,amsthm}
\usepackage{graphicx}
\usepackage[acronym]{glossaries}
\usepackage{xcolor}
\usepackage{comment}
\usepackage{cases}
\usepackage{subfigure}

\newacronym{siso}{SISO}{single-input single-output}
\newacronym{miso}{MISO}{multiple-input single-output}
\newacronym{mimo}{MIMO}{multiple-input multiple-output}
\newacronym{rf}{RF}{radio frequency}
\newacronym{awgn}{AWGN}{additive white Gaussian noise}
\newacronym{iid}{i.i.d.}{independent and identically distributed}
\newacronym{tx}{Tx}{transmitter}
\newacronym{rx}{Rx}{receiver}
\newacronym{milac}{MiLAC}{microwave linear analog computer}
\newacronym{snr}{SNR}{signal-to-noise ratio}
\newacronym{sim}{SIM}{stacked intelligent metasurface}
\newacronym{svd}{SVD}{singular value decomposition}
\newacronym{ris}{RIS}{reconfigurable intelligent surface}
\newacronym{bd-ris}{BD-RIS}{beyond diagonal RIS}
\newacronym{star-ris}{STAR-RIS}{simultaneously transmitting and reflecting RIS}
\newacronym{sre}{SRE}{smart radio environment}
\newacronym{los}{LoS}{line-of-sight}
\newacronym{nlos}{NLoS}{non-line-of-sight}
\newacronym{em}{EM}{electromagnetic}
\newacronym{fim}{FIM}{flexible intelligent metasurface}
\newacronym{fis}{FIS}{fixed intelligent surface}
\newacronym{bs}{BS}{base station}
\newacronym{irs}{IRS}{intelligent reflecting surface}
\newacronym{csi}{CSI}{channel state information}
\newacronym{ula}{ULA}{uniform linear array}
\newacronym{dft}{DFT}{discrete Fourier transform}
\newacronym{mrt}{MRT}{maximum ratio transmission}
\newacronym{egt}{EGT}{equal gain transmission}
\newacronym{mmwave}{mmWave}{millimeter wave}
\newacronym{thz}{THz}{terahertz}
\newacronym{ttd}{TTD}{true time delay}
\newacronym{fda}{FDA}{frequency diverse array}
\newacronym{fsa}{FSA}{frequency-switching array}
\newacronym{pls}{PLS}{physical-layer security}
\newacronym{film}{FILM}{flexible intelligent layered metasurface}

\DeclareMathOperator{\sign}{sign}

\begin{document}
\bstctlcite{BSTcontrol}

\title{Enabling Smart Radio Environments\\in the Frequency Domain With Movable Signals}

\author{Matteo~Nerini,~\IEEEmembership{Member,~IEEE}, and
        Bruno~Clerckx,~\IEEEmembership{Fellow,~IEEE}

\thanks{Corresponding author: Bruno Clerckx.}
\thanks{This work was supported in part by UKRI under Grant EP/Y004086/1, EP/X040569/1, EP/Y037197/1, EP/X04047X/1, EP/Y037243/1.}
\thanks{Matteo Nerini and Bruno Clerckx are with the Department of Electrical and Electronic Engineering, Imperial College London, SW7 2AZ London, U.K. (e-mail: m.nerini20@imperial.ac.uk; b.clerckx@imperial.ac.uk).}
\thanks{Bruno Clerckx is also with the Department of Electronic Engineering, Kyung Hee University, Yongin-si, Gyeonggi-do 17104, South Korea.}}

\maketitle

\begin{abstract}
\Glspl{sre} enhance wireless communications by allowing control over the channel.
They have been enabled through surfaces with reconfigurable \gls{em} properties, known as \glspl{ris}, and through flexible antennas, which can be viewed as realizations of \glspl{sre} in the \gls{em} domain and space domain, respectively.
However, these technologies rely on electronically reconfigurable or movable components, introducing implementation challenges that could hinder commercialization.
To overcome these challenges, we propose a new domain to enable \glspl{sre}, the frequency domain, through the concept of movable signals, where the signal spectrum can be dynamically moved along the frequency axis.
We first analyze movable signals in \gls{miso} systems under \gls{los} conditions, showing that they can achieve higher average received power than quantized \gls{egt}.
We then study movable signals under \gls{nlos} conditions, showing that they remain effective by leveraging reflections from surfaces made of uniformly spaced elements with fixed \gls{em} properties, denoted as \glspl{fis}.
Analytical results reveal that a \gls{fis}-aided system using movable signals can achieve up to four times the received power of a \gls{ris}-aided system using fixed-frequency signals.
\end{abstract}

\glsresetall


\section{Introduction}

Wireless communications have been advanced in recent years by the concept of \gls{sre}.
In a \gls{sre}, the wireless channel becomes an optimization variable that can be controlled to enhance performance and coverage \cite{dir19,dir20,wu20}.
Originally, the concept of \gls{sre} has been tightly linked to intelligent reflecting surface or \gls{ris} technology, which is widely regarded as its main enabler \cite{wu19,wu21}.
A \gls{ris} is a surface made of multiple passive elements with reconfigurable reflecting properties.
Thus, by controlling how a \gls{ris} reflects the incident signal, it is possible to steer it toward the intended destination, increasing the channel strength while mitigating interference.

\begin{figure}[t]
\centering
\includegraphics[width=0.32\textwidth]{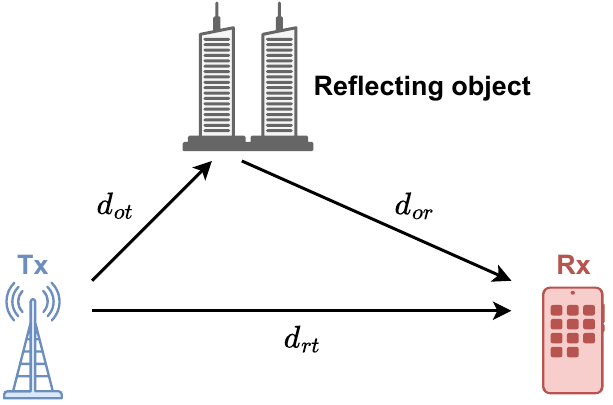}
\caption{Two-ray radio environment.}
\label{fig:toy}
\end{figure}

While \glspl{sre} have been predominantly associated with \gls{ris}, a broader perspective reveals that \gls{ris} is not the sole enabler of \glspl{sre}.
To identify other possible enablers, we can consider a simple channel model and analyze all the parameters impacting the wireless channel.
As the simplest multipath radio environment, consider the two-ray radio environment in Fig.~\ref{fig:toy}, where a single-antenna transmitter communicates with a single-antenna receiver through a two-ray channel.
This channel, denoted as $h$, can be modeled by the well-known two-ray model
\begin{equation}
h=\underbrace{\alpha_{rt}e^{-j2\pi\frac{fd_{rt}}{c}}}_{\text{LoS path}}+\underbrace{\alpha_{rot}e^{-j2\pi\frac{fd_{ro}}{c}}\Gamma e^{-j2\pi\frac{fd_{ot}}{c}}}_{\text{NLoS path}},\label{eq:h0}
\end{equation}
where the first additive term corresponds to the \gls{los} path and the second corresponds to the \gls{nlos} path reflected from a reflecting object in the environment.
In \eqref{eq:h0}, $c$ is the speed of light, $f$ is the frequency of the transmitted signal, $d_{rt}$, $d_{ro}$, and $d_{ot}$ are the distances between the transmitter and the receiver, the object and the receiver, and the transmitter and the object, respectively, and $\Gamma$ is the reflection coefficient of the reflecting object.
In addition, $\alpha_{rt}$ and $\alpha_{rot}$ are the path gains of the two paths, which are functions of the frequency and the distances.
From \eqref{eq:h0}, we observe that the channel $h$ depends on the three factors: the reflection coefficient $\Gamma$, the distances $d_{rt}$, $d_{ro}$, and $d_{ot}$, and the frequency $f$.
Thus, we identify three domains that can be exploited to reconfigure a wireless channel and enable a \gls{sre}:
\begin{enumerate}
\item \textit{\Gls{em} domain}, by acting on the \gls{em} properties of the antennas or the objects in the environment, hence acting on $\Gamma$ in \eqref{eq:h0};
\item \textit{Space domain}, by acting on the distances between the transmitting and the receiving antennas, and between them and the objects in the environment, hence acting on $d_{rt}$, $d_{ro}$, and $d_{ot}$ in \eqref{eq:h0};
\item \textit{Frequency domain}, by acting on the frequency of the transmitted signal, hence acting on $f$ in \eqref{eq:h0}.
\end{enumerate}

\glspl{sre} enabled in the \gls{em} domain have been extensively studied.
In this context, \gls{ris} has gathered significant attention for its potential to manipulate the wireless channel by reconfiguring the \gls{em} properties of reflecting surfaces deployed in the propagation environment \cite{wu19,wu21}.
To enhance the flexibility of \gls{ris} and extend the coverage, more general versions of \gls{ris} have been proposed.
Among them, \gls{star-ris} achieves full-space coverage by also allowing the transmission of the signal through its elements \cite{mu22}.
\Gls{bd-ris} generalizes conventional \gls{ris} by allowing interconnections between its elements, providing greater flexibility in shaping the radio environment \cite{she22}.
Additionally, \gls{sim} utilizes a multi-layer design, where multiple metasurface layers are cascaded, allowing for more complex analog-domain processing \cite{an23}.

\begin{figure}[t]
\centering
\includegraphics[width=0.42\textwidth]{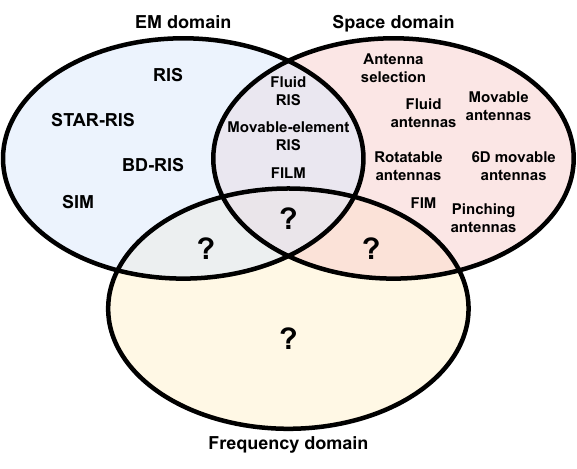}
\caption{Current enablers of SREs in the EM, space, and frequency domains.}
\label{fig:enablers}
\end{figure}

\glspl{sre} have also been enabled in the space domain by physically moving the transmitting or receiving antennas, hence effectively optimizing the distances between them and the objects in the environment.
The simplest way to enable space-domain \glspl{sre} is through antenna selection, a classical technique in \gls{mimo} communications that uses \gls{rf} switches to activate the antennas experiencing the best channels \cite{mol04}.
More recently, several flexible antenna technologies have been proposed, such as fluid antennas \cite{won21,wu25}, movable antennas \cite{zhu24-1}, rotatable antennas \cite{zhe25}, and 6D movable antennas \cite{sha25}.
Different from fluid and movable antennas, in \glspl{fim}, the antenna elements are located on a morphing surface able to reconfigure its 3D shape \cite{an25}.
While these flexible antenna technologies involve moving antennas by distances comparable to the wavelength, pinching antennas is another interesting solution that utilizes waveguides to offer greater movement flexibility \cite{din25}.
Further works have investigated \glspl{sre} enabled jointly in the \gls{em} and space domains.
In this direction, fluid \gls{ris} \cite{sal25} and movable-element \gls{ris} \cite{hu24} have been proposed as more flexible \glspl{ris} whose elements have a reconfigurable location.
\Gls{film} has also emerged by combining the strength of \gls{sim} and \gls{fim} \cite{niu25}.
In Fig.~\ref{fig:enablers}, the enablers of \glspl{sre} in the three domains are visualized in a Venn diagram, where the question marks mean that enabling \glspl{sre} in the frequency domain remains less explored.

Although \glspl{sre} in the \gls{em} and space domains have been widely investigated in previous works \cite{wu19}-\cite{niu25}, the potential of enabling \glspl{sre} in the frequency domain remains less explored.
To fill this gap, in this paper, we show that \glspl{sre} can be enabled also in the frequency domain by reconfiguring the frequency of the transmitted signal.
We refer to this technique as \textit{movable signals}, since the signal spectrum can be ``moved'' along the frequency axis.
To obtain significant performance gains with movable signals even when the \gls{los} link between the transmitter and receiver is obstructed, we leverage surfaces composed of uniformly spaced elements with fixed reflection properties deployed in the propagation environment, which we refer to as \textit{\glspl{fis}}.\footnote{The acronym \gls{fis} is chosen to highlight the duality with \gls{ris}.
The ``intelligence'' in \gls{fis} manifests in their offline design (e.g., including \gls{em} properties and geometry) rather than in their online reconfiguration as in \gls{ris}.}

The difference between frequency-domain \glspl{sre} and the widely explored \gls{em}- and space-domain \glspl{sre} can be intuitively stated as follows.
\textit{While with \gls{ris} we reconfigure the \gls{em} properties of the scattering objects, in frequency-domain \glspl{sre} we reconfigure the frequency of the transmitted signal and exploit the natural frequency selectivity of those \gls{em} properties.
Besides, while with flexible antennas we move the antennas in space and optimize the distances, in frequency-domain \glspl{sre} we ``move'' the signal in frequency and optimize its wavelength.}
Thus, frequency-domain \glspl{sre} offer a key advantage over \gls{em}- and space-domain \glspl{sre} by eliminating the need for electronically or mechanically tunable components, as required in \gls{ris} and flexible antennas.
The trade-off is that frequency-domain \glspl{sre} require access to a wide frequency range, necessary to optimize the signal frequency with sufficient flexibility and meaningfully control the channel.

A straightforward way to realize frequency-domain \glspl{sre} is through opportunistic frequency selection, a technique commonly used by operators when scheduling users.
However, the available frequency spectrum is typically divided into a limited number of closely spaced slots, and small variations in frequency have only a minor impact on the channel in \gls{los} scenarios (which are expected to be dominant in future communications).
Unlike conventional systems, where frequency selection happens within a relatively narrow range, movable signals treat frequency as a variable reconfigured within a much wider range.
This flexibility transforms the frequency into a powerful ``knob'' to perform channel shaping and enable \glspl{sre}, resulting in novel optimization problems.
To enable movable signals to ``move'' over a sufficiently wide frequency range under limited spectrum resources, an intelligent allocation of the available spectrum among operators is required.
For instance, it will be more beneficial to have access to multiple spectrum fragments distributed across a wide frequency range and to optimize the carrier frequency of movable signals within those fragments, rather than selecting a frequency slot among closely spaced ones.
Mechanisms such as inter-band carrier aggregation, already available in 4G and 5G systems, could be leveraged to practically realize movable signals \cite{she12,han23}.

The idea of improving the communication link (or optimizing the radiation pattern) by controlling the operating frequency dates back to the 1940s, when leaky-wave antennas were proposed \cite[Chapter~10]{joh93}, \cite{jac12}.
A leaky-wave antenna employs a waveguide along which the \gls{em} signal propagates and is gradually radiated, or ``leaked''.
An interesting property of this structure is that the beam angle varies with the signal frequency.
Similar principles are used in frequency scanning antennas, which emerged in the 1950s \cite[Chapter~19]{joh93}, \cite{ish62}.
These concepts have been proposed to enable frequency-dependent beam steering in \gls{thz} communications, as demonstrated for leaky-wave antennas in \cite{kar15,mur17,gha20,gab23,zhu25} and for frequency-scanning antennas in \cite{cul12,sar18}.
In addition, recent inspiring work \cite{zha20,zha21} has proposed integrating \gls{ttd} devices into phased arrays to steer the subcarriers of a wideband \gls{thz} signal toward different directions.
Arrays equipped with \gls{ttd} devices have also been used for \gls{mmwave} systems, to create parallel frequency-dependent training beams and therefore achieve fast beam training \cite{yan19,bol21} and low-latency multiple access \cite{li22}, as comprehensively reviewed in \cite{lin22}.
Furthermore, it has been observed that introducing small frequency offsets across the antennas of a so-called \gls{fda} results in range-dependent beamforming \cite{wan15}.
To improve \glspl{fda}, \glspl{fsa} have been proposed, where the carrier frequency can also be reconfigured jointly with the per-antenna frequency offsets \cite{zho25}.
Remarkably, \glspl{fsa} enable highly flexible beam control, comparable to that of movable antennas, for enhancing \gls{pls} in \gls{thz} communications.
Reconfiguring the transmission frequency is therefore expected to play a crucial role in \gls{pls} \cite{niu26}.

Building on previous work on frequency-dependent beamforming, movable signals offer broader opportunities for enhancing wireless communications.
There are four differences between movable signals and frequency-dependent beamforming solutions proposed in previous literature.
First, movable signals can be employed independently of the antenna array architecture.
They are also beneficial when the transmitter is simply equipped with a linear array of isotropic antennas, as we will demonstrate in Section~\ref{sec:fundamentals-los}, or even a single isotropic antenna, as shown in Section~\ref{sec:fundamentals-nlos}.
Second, unlike beam scanning solutions which primarily benefit \gls{los} scenarios, movable signals are also effective in \gls{nlos} by exploiting the reflections from non-reconfigurable reflecting surfaces, as we will show in Section~\ref{sec:fundamentals-nlos}.
In \gls{nlos}, movable signals can also be combined with a fixed precoder to enable a form of opportunistic beamforming \cite{vis02}, where frequency-dependent channel fluctuations are exploited to select favorable propagation conditions.
Third, movable signals can be integrated with existing digital or analog beamforming techniques.
By jointly optimizing both the signal precoding, as conventionally done, and the signal frequency, movable signals can further enhance flexibility and performance, as we will discuss in Section~\ref{sec:performance}.
Fourth, unlike \glspl{fda} and \glspl{fsa}, movable signals do not require a frequency offset at each antenna in the case of a multi-antenna transmitter.
The contributions of this paper are the following.

\textit{First}, we propose frequency-domain \glspl{sre} as a novel type of \glspl{sre} enabled by dynamically tuning the center frequency of the transmitted signal, an approach referred to as movable signals.
Movable signals allow shaping the wireless channel by acting solely on the signal frequency.
Thus, they eliminate the need for electronically or mechanically tunable hardware components, which are used in widely investigated \gls{em}- and space-domain \glspl{sre}.

\textit{Second}, we show that movable signals can maximize the received power in \gls{miso} systems under \gls{los} conditions in Section~\ref{sec:fundamentals-los}.
To this end, we provide a closed-form expression for the optimal frequency, characterizing the maximum received power and the region of space where such a maximum received power can be achieved, namely the coverage.
Movable signals can maximize the received power without the need for reconfigurable \gls{rf} phase shifters, which are commonly required in \gls{egt}, hence reducing ohmic losses and hardware complexity.

\textit{Third}, we demonstrate that movable signals remain effective in \gls{nlos} conditions by exploiting the reflections from surfaces with fixed (non-reconfigurable) reflecting properties, referred to as \glspl{fis}.
In Section~\ref{sec:fundamentals-nlos}, considering \gls{nlos} systems aided by \gls{fis}, we derive the optimal frequency of movable signals that maximizes the received power, and investigate the fundamental limits of the received power and coverage.
Interestingly, movable signals used in a \gls{fis}-aided system can achieve up to four times the received power of a conventional \gls{ris}-aided system where the signal frequency is not reconfigured.
At the same time, the hardware requirements are significantly simplified since a \gls{fis} is non-reconfigurable.

\textit{Fourth}, we propose a practical transmission protocol to optimize the frequency of movable signals in both \gls{los} and \gls{nlos}, and compare the achieved performance with quantized \gls{egt} and \gls{ris}, respectively.
Numerical results in Section~\ref{sec:performance} show that movable signals can offer a significantly higher received power at the cost of a reduced coverage compared to quantized \gls{egt} and \gls{ris}.
On average, movable signals obtain a higher received power than quantized \gls{egt} and \gls{ris}, remarkably without requiring reconfigurable \gls{rf} components, such as phase shifters and tunable loads.
Considerable performance can already be obtained when movable signals (assumed to be narrowband) can move their carrier frequency within a range approximately equal to $10\%$ of the carrier frequency.
As expected, the performance of movable signals improves as the width of the available frequency range increases.


\textit{Notation}:
Vectors and matrices are denoted with bold lower and bold upper letters, respectively.
Scalars are represented with letters not in bold font.
$\Re(a)$, $\Im(a)$, $\vert a\vert$, and $\arg(a)$ refer to the real part, imaginary part, absolute value, and phase of a complex scalar $a$, respectively.
$\mathbf{a}^T$, $\mathbf{a}^H$, $[\mathbf{a}]_{i}$, and $\Vert\mathbf{a}\Vert$ refer to the transpose, conjugate transpose, $i$th element, and $\ell_2$-norm of a vector $\mathbf{a}$, respectively.
$[\mathbf{A}]_{i,k}$ and $\Vert\mathbf{A}\Vert$ refer to the $(i,k)$th element and $\ell_2$-norm (or spectral norm) of a matrix $\mathbf{A}$, respectively.
$\mathbb{Z}$ and $\mathbb{C}$ denote the integer and complex number sets, respectively.
$j=\sqrt{-1}$ denotes the imaginary unit.
$\mathbf{I}$, $\mathbf{0}$, and $\mathbf{1}$ denote the identity matrix, the all-zero matrix, and the all-one matrix, respectively.
Given a real scalar $a$, $\sign(a)$ is the sign function defined as $\sign(a)=-1$ if $a<0$, $\sign(a)=1$ if $a>0$, and $\sign(a)=0$ if $a=0$.

\section{Enabling Frequency-Domain SREs in LoS\\with Movable Signals}
\label{sec:fundamentals-los}

In this section, we introduce the fundamentals of \glspl{sre} enabled in the frequency domain through movable signals.
Considering a single-user \gls{miso} system in \gls{los}, we show that movable signals can be used as an alternative to \gls{egt} to maximize the received signal power with highly reduced hardware complexity.

\subsection{System Model}

\begin{figure}[t]
\centering
\includegraphics[width=0.36\textwidth]{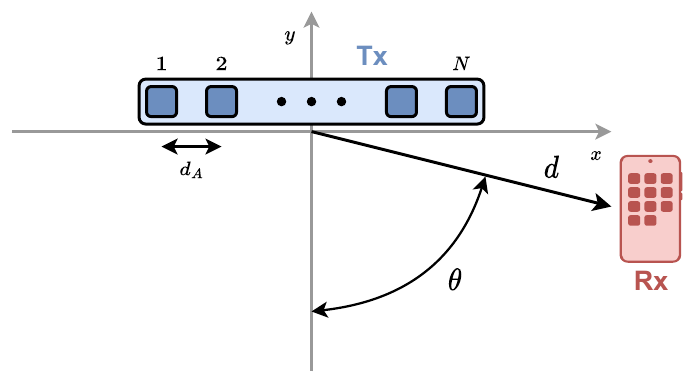}
\caption{LoS communication between a multi-antenna transmitter and a single-antenna receiver.}
\label{fig:system-los}
\end{figure}

Consider the \gls{miso} system represented in Fig.~\ref{fig:system-los}, where a multi-antenna transmitter equipped with $N$ antennas communicates with a single-antenna receiver in \gls{los}.
The \gls{los} propagation is considered as it is expected to be essential in future networks, especially operating at \gls{mmwave} and \gls{thz} frequencies.
The transmitted signal $\mathbf{x}\in\mathbb{C}^{N\times 1}$ is given by $\mathbf{x}=\mathbf{w}s$, as a function of the transmitted symbol $s\in\mathbb{C}$ and the precoder $\mathbf{w}\in\mathbb{C}^{N\times 1}$.
The transmitted symbol $s$ is such that $\mathbb{E}[\vert s\vert^2]=P_T$, where $P_T$ is the transmitted signal power, and the precoder $\mathbf{w}$ satisfies $\Vert\mathbf{w}\Vert^2=1$.
Thus, the received signal $y\in\mathbb{C}$ writes as $y=\mathbf{h}\mathbf{x}+n$, where $\mathbf{h}\in\mathbb{C}^{1\times N}$ is the wireless channel between the transmitter and the receiver and $n\in\mathbb{C}$ is the noise.

To model the channel $\mathbf{h}$, we assume the transmitting array to be a \gls{ula} located along the $x$-axis and centered in $x=0$, such that the $n$th transmitting antenna has $x$ coordinate
\begin{equation}
x_n=\left(n-\frac{N+1}{2}\right)d_A,\label{eq:xn}
\end{equation}
where $d_A$ is the antenna spacing.
Denoting as $d$ the distance between the receiver and the center of the transmitting array, and as $\theta\in[-\pi/2,\pi/2]$ the angle of the receiver direction with respect to the transmitting array normal, the distance between the receiver and the $n$th transmitting antenna is
\begin{equation}
d_n=d-x_n\sin\left(\theta\right),\label{eq:dn}
\end{equation}
assuming far-field propagation.
Therefore, the $n$th entry of the channel $\mathbf{h}$ writes as
\begin{equation}
\left[\mathbf{h}\right]_n
=e^{-j\frac{2\pi}{\lambda}d_{n}}
=e^{-j\frac{2\pi}{\lambda}\left[d-\left(n-\frac{N+1}{2}\right)d_A\sin(\theta)\right]},\label{eq:h-los}
\end{equation}
for $n=1,\ldots,N$, where $\lambda=\frac{c}{f}$ is the wavelength, $c$ the speed of light, and $f$ the frequency, as a consequence of \eqref{eq:xn} and \eqref{eq:dn}.
Note that this system model includes several standard assumptions commonly considered in related literature, i.e., the antennas are assumed to be perfectly matched isotropic radiators and mutual coupling effects between them are assumed to be negligible (which is a good approximation when the antenna spacing $d_A$ is at least half-wavelength).

\subsection{Received Signal Power}

In this \gls{miso} system, the goal is to maximize is the received signal power, given by
\begin{equation}
P_R=P_T\left\vert\mathbf{h}\mathbf{w}\right\vert^2,\label{eq:pr1-los}
\end{equation}
which is a function of the channel $\mathbf{h}$ and the precoder $\mathbf{w}$.
Although other metrics can be considered, such as the achievable rate $R=\log_2(1+P_R/\sigma^2)$, where $\sigma^2$ is the noise power, or the outage probability $P_{out}=\text{Pr}\{R\leq \bar{R}\}$, where $\bar{R}$ is the target rate, they are maximized when the received signal power is maximized, and we can therefore focus purely on this metric.
Conventionally, the received signal power $P_R$ is maximized by optimizing the precoder $\mathbf{w}$ as a function of the channel $\mathbf{h}$, which is assumed to be known and fixed.
It is well-known that the optimal solution for the precoder $\mathbf{w}$ is given by \gls{mrt}, which boils down to \gls{egt} in the case of \gls{los} channels.
However, there are two methods to implement a reconfigurable precoder $\mathbf{w}$, both problematic as the number of antennas $N$ grows large.
First, $\mathbf{w}$ can be implemented with digital beamforming, i.e., the signal $\mathbf{x}=\mathbf{w}s$ is computed in baseband and then converted to \gls{rf} through $N$ \gls{rf} chains, each connected to one of the $N$ transmitting antennas.
This method is not scalable to high numbers of antennas due to the high cost and power consumption of the \gls{rf} chains.
Second, $\mathbf{w}$ can be implemented with analog beamforming, where only one \gls{rf} chain is needed to carry the symbol $s$, and $N$ analog phase shifters connected to the $N$ transmitting antennas adjust the phase of the signals at the antennas.
This second method suffers from a performance degradation due to the significant insertion losses of reconfigurable phase shifters.
To radically address these problems arising when reconfiguring the precoder $\mathbf{w}$, we aim at maximizing the received signal power $P_R$ by reconfiguring the channel $\mathbf{h}$ while maintaining $\mathbf{w}$ fixed.
To this end, we propose to reconfigure $\mathbf{h}$ given in \eqref{eq:h-los} by optimizing the wavelength $\lambda$, or, equivalently, the frequency $f$ of the transmitted signal, effectively enabling a \gls{sre} in the frequency domain.

\begin{figure}[t]
\centering
\includegraphics[width=0.32\textwidth]{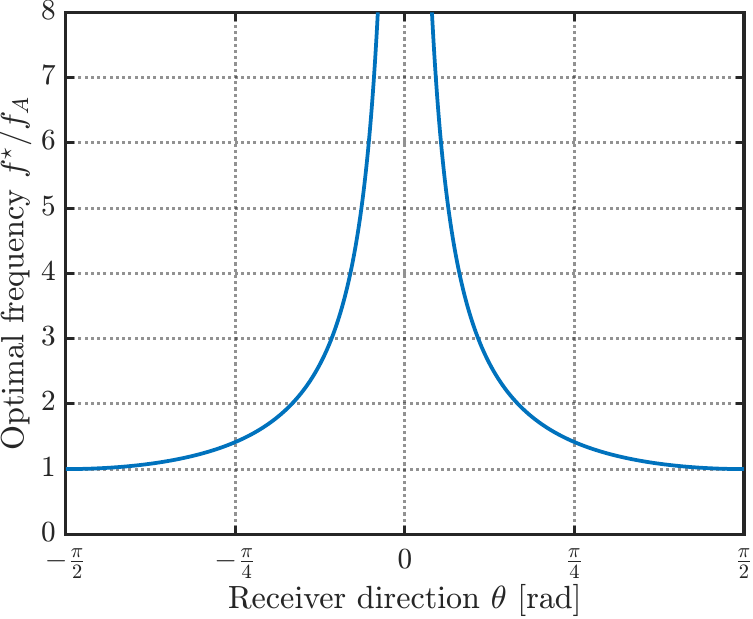}
\caption{Optimal frequency $f^\star$ normalized by $f_A=c/d_A$.}
\label{fig:f-los}
\end{figure}

\begin{figure*}[t]
\centering
\subfigure[]{
\includegraphics[width=0.32\textwidth]{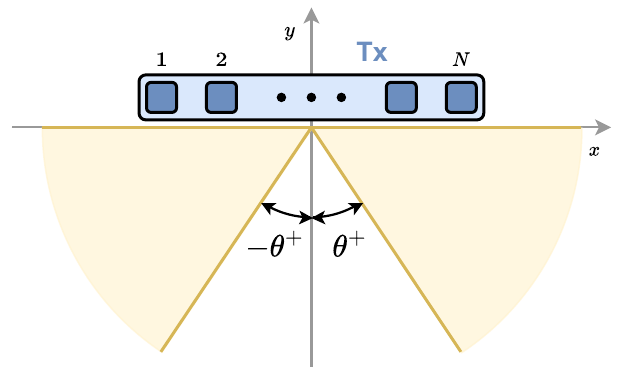}
\label{fig:system-los-coverage-a}}
\subfigure[]{
\includegraphics[width=0.42\textwidth]{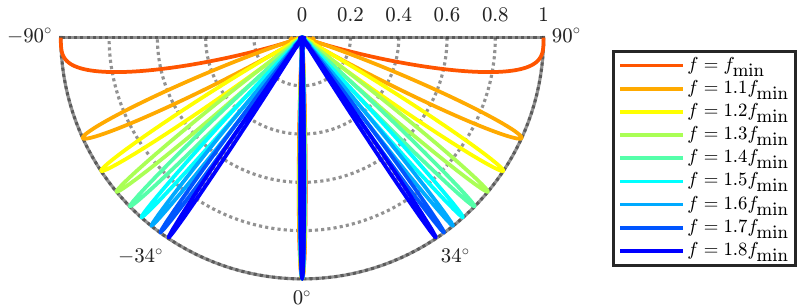}
\label{fig:system-los-coverage-b}}
\caption{Multi-antenna transmitter using movable signals with frequency range width $W=1.8$, having $\theta^+=34^\circ$ according to \eqref{eq:t+los}.
(a) The region within coverage is highlighted in yellow.
(b) The radiation pattern is shown for different frequencies $f$.}
\end{figure*}

Our problem is therefore to maximize the received signal power $P_R$ by optimizing the wavelength $\lambda$ on a per-channel realization basis, i.e.,
\begin{equation}
\underset{\lambda}{\mathsf{\mathrm{max}}}\;\;
P_T\left\vert\mathbf{h}\mathbf{w}\right\vert^2\;\;
\mathsf{\mathrm{s.t.}}\;\;
\eqref{eq:h-los},\;\mathbf{w}\text{ is fixed},\label{eq:prob-los}
\end{equation}
where $\mathbf{w}$ is optimized offline and therefore remains fixed during the deployment.
Interestingly, problem \eqref{eq:prob-los} can be solved with a global optimal solution available in closed form.
To derive it, we first derive an upper bound on the objective function $P_R$ and then provide expressions for a fixed $\mathbf{w}$ and a reconfigurable $\lambda$ which are proven to achieve it.
First, the received signal power is upper bounded by
\begin{equation}
P_R
\leq P_T\left\Vert\mathbf{h}\right\Vert^2\left\Vert\mathbf{w}\right\Vert^2
=P_TN,\label{eq:ub-los}
\end{equation}
where the inequality follows from the Cauchy-Schwarz inequality and the equality holds since $\Vert\mathbf{h}\Vert^2=N$ and $\Vert\mathbf{w}\Vert^2=1$.
Thus, to achieve the upper bound $P_R^\star=P_TN$, we fix $\mathbf{w}=\mathbf{1}/\sqrt{N}$ and show that $\lambda$ can be optimally reconfigured on a per-channel realization basis.

With $\mathbf{w}=\mathbf{1}/\sqrt{N}$ and $\mathbf{h}$ given by \eqref{eq:h-los}, the received signal power becomes
\begin{equation}
P_R
=\frac{P_T}{N}\left|\sum_{n=1}^Ne^{-j\frac{2\pi}{\lambda}\left[d-\left(n-\frac{N+1}{2}\right)d_A\sin(\theta)\right]}\right|^2,\label{eq:pr2-los}
\end{equation}
indicating that we achieve the upper bound $P_R^\star=P_TN$ if and only if all the $N$ additive terms in \eqref{eq:pr2-los} are co-phased, i.e., they have the same phase up to a $2\pi$ factor.
Such an optimality condition can be formulated as
\begin{equation}
-\frac{2\pi}{\lambda}\left[d-\left(n-\frac{N+1}{2}\right)d_A\sin(\theta)\right]=\alpha+K_n2\pi,\label{eq:cond1-los}
\end{equation}
for $n=1,\ldots,N$, where $\alpha\in[0,2\pi)$ is an arbitrary phase and $K_n\in\mathbb{Z}$ is an integer number that can vary with the index $n$.
By considering
\begin{equation}
\alpha=-\frac{2\pi}{\lambda}\left(d+\frac{N+1}{2}d_A\sin\left(\theta\right)\right),
\end{equation}
condition \eqref{eq:cond1-los} simplifies as
\begin{equation}
\frac{n}{\lambda}d_A\sin(\theta)=K_n,\label{eq:cond2-los}
\end{equation}
for $n=1,\ldots,N$.
Note that \eqref{eq:cond2-los} is satisfied for any $\lambda$ when $\theta=0$, since an array with $\mathbf{w}=\mathbf{1}/\sqrt{N}$ transmits in the broadside direction $\theta=0$ at any frequency.
For $\theta\neq0$, \eqref{eq:cond2-los} is satisfied by setting the wavelength as a function of $\theta$ as
$\lambda^\star(\theta)=nd_A\sin(\theta)/K_n$,
indicating that there are infinitely many possible values of $\lambda^\star$, depending on the values $K_n$.
Among all the possible values of $\lambda^\star$, we select the largest one, since it corresponds to the lowest frequency $f^\star=c/\lambda^\star$, which is obtained by considering $K_n=n\sign(\sin(\theta))$, for $n=1,\ldots,N$, and is given by
$\lambda^\star(\theta)=d_A\vert\sin(\theta)\vert$,
having a corresponding frequency $f^\star=c/\lambda^\star$ given by
\begin{equation}
f^\star\left(\theta\right)=\frac{f_A}{\left\vert\sin(\theta)\right\vert},\label{eq:f-los}
\end{equation}
where we have introduced $f_A=c/d_A$ as the frequency with wavelength equal to the antenna spacing of the transmitter.

We have shown that the received signal power $P_R$ can be maximized by fixing the precoder to $\mathbf{w}=\mathbf{1}/\sqrt{N}$ and controlling the channel $\mathbf{h}$ through movable signals, i.e., signals with reconfigurable frequency.
By optimizing the signal frequency $f$ based on the receiver direction $\theta$, as defined in \eqref{eq:f-los} and illustrated in Fig.~\ref{fig:f-los}, the maximum received signal power $P_R^\star=P_TN$ can be achieved.
This maximum performance is the same as the maximum performance of \gls{egt} with $[\mathbf{w}^\star]_n=e^{-j\arg([\mathbf{h}]_n)}/\sqrt{N}$, for $n=1,\ldots,N$, assuming that it is realized via digital beamforming or with analog phase shifters that are lossless and arbitrarily reconfigurable.
At the same time, movable signals only require the \gls{rf} chain to be connected to the antennas through a power divider, with no reconfigurable \gls{rf} components such as phase shifters, significantly reducing hardware complexity, power consumption, and ohmic losses.
As a trade-off, the carrier frequency of movable signals must be adjusted over a wide frequency range to cover all possible directions $\theta$.
In the next subsection, we examine how wide this frequency range needs to be to obtain satisfactory performance.

\subsection{Coverage}

The optimal frequency $f^\star$ in \eqref{eq:f-los} could take any positive real value depending on $\theta$, as shown in Fig.~\ref{fig:f-los}.
However, $f^\star$ cannot be reconfigured to any value in practice since only a limited frequency range is available.
It is expected that $f^\star$ can be selected in practice within a range $[f_{\textrm{min}},f_{\textrm{max}}]$, where we write $f_{\textrm{max}}=Wf_{\textrm{min}}$, with $W>1$ being the width of the frequency range.\footnote{If, for example, $f_{\textrm{min}}=10$~GHz and $W=1.2$, the transmission is allowed within the frequency range 10-12~GHz.
However, this does not imply that the signal occupies the whole 2~GHz bandwidth.
Instead, the transmission occurs in a subchannel centered at $f^\star$ located within this 2~GHz range.}
With this constraint on the operating frequency, the received signal power upper bound $P_R^\star=P_TN$ is achievable only for specific values of $\theta$, namely in a limited region of space.
In the following, we characterize the size of the region of space where it is possible to achieve $P_R^\star=P_TN$ with a frequency constrained in the range $[f_{\textrm{min}},f_{\textrm{max}}]$, which we denote as the coverage of a frequency-domain \gls{sre} enabled by movable signals.

To achieve maximum coverage, we set $f_{\textrm{min}}$ as the minimum optimal frequency $f^\star$ given by \eqref{eq:f-los}, which occurs in $\theta=\pm\pi/2$ and is given by
\begin{equation}
f_{\textrm{min}}=f^\star\left(\pm\pi/2\right)=f_A.\label{eq:f-min-los}
\end{equation}
Thus, noticing that $f^\star$ in \eqref{eq:f-los} is a symmetric function of $\theta$ increasing in $[-\pi/2,0)$, tending to $+\infty$ as $\theta\to0$, and decreasing in $(0,\pi/2]$, there exists a value of $\theta$ in $(0,\pi/2]$, denoted as $\theta^+$, such that $f^\star(-\theta^+)=f^\star(\theta^+)=f_{\textrm{max}}$, i.e.,
\begin{equation}
f^\star\left(\theta^+\right)=\frac{f_A}{\sin(\theta^+)}=f_{\textrm{max}},\label{eq:f-max-los}
\end{equation}
following \eqref{eq:f-los}.
By inverting \eqref{eq:f-max-los}, such a $\theta^+$ is obtained as
\begin{equation}
\theta^+
=\arcsin\left(\frac{f_A}{f_{\textrm{max}}}\right)
=\arcsin\left(\frac{1}{W}\right),\label{eq:t+los}
\end{equation}
where we exploited $f_{\textrm{max}}=Wf_{\textrm{min}}$, with $f_{\textrm{min}}=f_A$ to write $\theta^+$ purely as a function of the frequency range width $W$.
Since $f^\star$ in \eqref{eq:f-los} is symmetric, increasing in $[-\pi/2,0)$, and decreasing in $(0,\pi/2]$, we have that for any receiver direction $\theta\in[-\pi/2,-\theta^+]\cup[\theta^+,\pi/2]$ it holds $f^\star\in[f_{\textrm{min}},f_{\textrm{max}}]$ and the performance upper bound $P_R^\star=P_TN$ can be achieved.

\begin{figure}[t]
\centering
\includegraphics[width=0.32\textwidth]{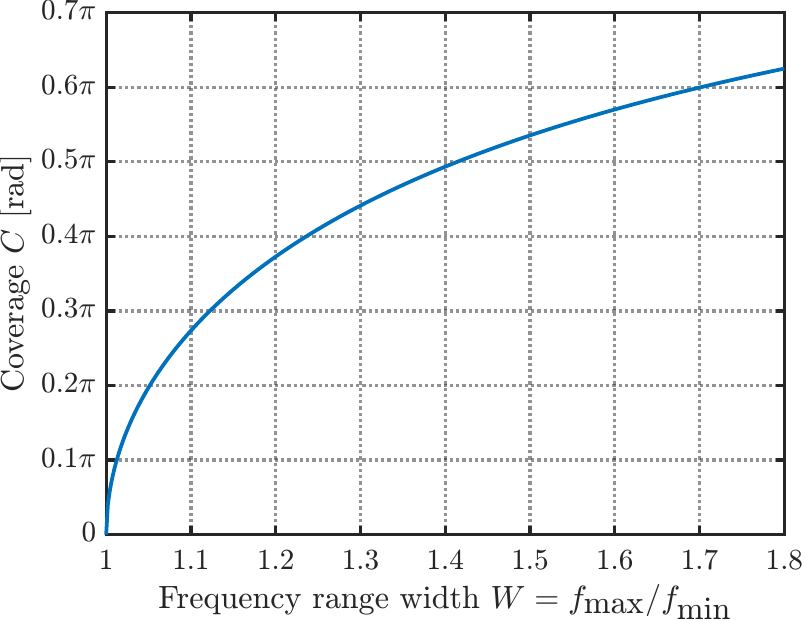}
\caption{Coverage $C$ versus the frequency range width $W=f_{\text{max}}/f_{\text{min}}$.}
\label{fig:coverage-los}
\end{figure}

In summary, the upper bound $P_R^\star=P_TN$ can be achieved in the region of space where $\theta\in[-\pi/2,-\theta^+]\cup[\theta^+,\pi/2]$, as highlighted in yellow in Fig.~\ref{fig:system-los-coverage-a}, and the coverage of movable signals is
\begin{equation}
C
=\left(-\theta^++\pi/2\right)+\left(\pi/2-\theta^+\right)
=\pi-2\theta^+,\label{eq:C}
\end{equation}
where $\theta^+$ is given by \eqref{eq:t+los} as a function of the frequency range width $W$.
In Fig.~\ref{fig:system-los-coverage-b}, the radiation pattern of the transmitter is shown for different frequencies $f\in[f_{\textrm{min}},f_{\textrm{max}}]$, computed as $R(\theta)=\vert\mathbf{a}(\theta)\mathbf{w}\vert^2$, where $\mathbf{a}(\theta)=[1,e^{j\frac{2\pi}{\lambda}d_A\sin(\theta)},\ldots,e^{j\frac{2\pi}{\lambda}d_A(N-1)\sin(\theta)}]/\sqrt{N}$ is the steering vector and $\mathbf{w}=\mathbf{1}/\sqrt{N}$, with $N=16$.
We observe that a beam at $\theta=0$ is always present for any frequency $f$, while the receiver is served through one of the two grating lobes, whose direction depend on $f$.\footnote{If the frequency can take only a limited number of values within the range $[f_{\textrm{min}},f_{\textrm{max}}]$, only restricted set of beam directions can be covered, as shown in Fig.~\ref{fig:system-los-coverage-b} where nine frequencies are considered.
In this case, the available frequency values should be carefully designed to avoid blind spots, i.e., directions that cannot be effectively served, which is an interesting direction for future research.}

We report the coverage $C$ as a function of $W$ in Fig.~\ref{fig:coverage-los}, where we observe that the coverage increases with $W$, since a larger $W$ implies additional flexibility in choosing the optimal frequency.
This indicates that we can achieve half-space coverage, i.e., $C=\pi$, only with an infinite frequency range available, while in practice the coverage is reduced, i.e., $C<\pi$.
Nevertheless, significant coverage can also be achieved with a frequency range width $W\leq 1.8$.
The choice of the operating point along this trade-off between coverage and frequency range width depends on practical deployment considerations and objectives.
On the one hand, operating with a smaller frequency range is more practical, but results in reduced coverage, making it suitable for scenarios where users are geographically concentrated or where multiple \glspl{fis} can be deployed to cover smaller areas.
On the other hand, a wider frequency range is necessary to optimally serve users distributed over a larger area.

\section{Enabling Frequency-Domain SREs in NLoS\\with Movable Signals\\and Fixed Intelligent Surfaces (FISs)}
\label{sec:fundamentals-nlos}

We have introduced the fundamentals of frequency-domain \glspl{sre} enabled with movable signals assuming \gls{los} conditions.
We have shown that movable signals can maximize the received signal power in \gls{los} systems, offering an alternative to \gls{egt} with reduced hardware complexity.
In this section, we characterize the fundamental limits of frequency-domain \glspl{sre} in \gls{nlos} systems.
When the \gls{los} link between the transmitter and receiver is obstructed, movable signals can still be effective by leveraging reflections from surfaces with fixed reflecting properties, referred to as \glspl{fis}.

\subsection{System Model}

\begin{figure}[t]
\centering
\includegraphics[width=0.36\textwidth]{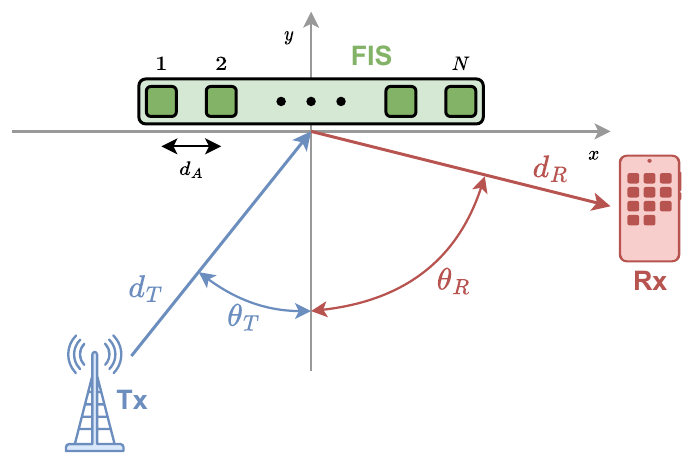}
\caption{NLoS communication between a single-antenna transmitter and a single-antenna receiver aided by a FIS.}
\label{fig:system-nlos}
\end{figure}

Consider a \gls{siso} system between a single-antenna transmitter and a single-antenna receiver, where the \gls{los} link is obstructed.
To enhance the channel strength in this \gls{nlos} scenario, we assume that an $N$-element surface with fixed \gls{em} properties has been deployed in the propagation environment, which we denote as \gls{fis}.
This \gls{fis} is suitably located in \gls{los} with both the transmitter and receiver, as represented in Fig.~\ref{fig:system-nlos}.
Denoting as $x\in\mathbb{C}$ the transmitted signal, such that $\mathbb{E}[\vert x\vert^2]=P_T$, where $P_T$ is the transmitted signal power, the received signal $y\in\mathbb{C}$ writes as $y=hx+n$, where $h\in\mathbb{C}$ is the wireless channel and $n\in\mathbb{C}$ is the noise.

Following previous literature on \gls{ris} \cite{wu19,wu21}, the channel $h$ can be expressed as
\begin{equation}
h
=\mathbf{h}_{R}\boldsymbol{\Theta}\mathbf{h}_{T}-\mathbf{h}_{R}\mathbf{h}_{T},\label{eq:h1-nlos}
\end{equation}
where $\boldsymbol{\Theta}\in\mathbb{C}^{N\times N}$ is a diagonal matrix $\boldsymbol{\Theta}=\text{diag}(e^{j\theta_1},\ldots,e^{j\theta_N})$, with $e^{j\theta_n}$ being the reflection coefficient of the $n$th \gls{fis} element, $\mathbf{h}_{R}\in\mathbb{C}^{1\times N}$ is the channel between the \gls{fis} and receiver, and $\mathbf{h}_{T}\in\mathbb{C}^{N\times 1}$ is the channel between the transmitter and \gls{fis}.
In \eqref{eq:h1-nlos}, the direct link is assumed to be obstructed and we account for the specular reflection of the surface via the term $-\mathbf{h}_{R}\mathbf{h}_{T}$, which is due to the structural scattering component of the surface \cite{nos24,abr24,ner24}.

To model $\mathbf{h}_{R}$ and $\mathbf{h}_{T}$, we assume for simplicity the \gls{fis} to be a \gls{ula} in \gls{los} with the transmitter and receiver.
Considering the \gls{fis} elements to be located along the $x$-axis and the \gls{fis} to be centered in $x=0$, the $n$th element has $x$ coordinate $x_n$ as given by \eqref{eq:xn}, where $d_A$ is the antenna spacing of the \gls{fis}.
Denoting as $d_{R}$ the distance between the receiver and the center of the \gls{fis}, and as $\theta_R\in[-\pi/2,\pi/2]$ the angle of the receiver direction with respect to the \gls{fis} normal, the distance between the receiver and the $n$th \gls{fis} element is given by $d_{R,n}=d_{R}-x_n\sin(\theta_R)$ in the far-field.
Thus, the entries of the channel $\mathbf{h}_{R}$ write as
\begin{equation}
\left[\mathbf{h}_{R}\right]_n
=e^{-j\frac{2\pi}{\lambda}d_{R,n}}
=e^{-j\frac{2\pi}{\lambda}\left[d_{R}-\left(n-\frac{N+1}{2}\right)d_A\sin(\theta_R)\right]},\label{eq:hR}
\end{equation}
for $n=1,\ldots,N$.
Similarly, defining $d_{T}$ the distance between the transmitter and the center of the \gls{fis}, and as $\theta_T\in[-\pi/2,\pi/2]$ the angle of the transmitter direction with respect to the \gls{fis} normal, the entries of $\mathbf{h}_{T}$ write as
\begin{equation}
\left[\mathbf{h}_{T}\right]_n=e^{-j\frac{2\pi}{\lambda}\left[d_{T}-\left(n-\frac{N+1}{2}\right)d_A\sin(\theta_T)\right]},\label{eq:hT}
\end{equation}
for $n=1,\ldots,N$.
Several standard assumptions are considered in this system model, i.e., the \gls{fis} elements are assumed to be perfectly matched isotropic radiators with negligible mutual coupling effects, which is reasonable with antenna spacing larger than half-wavelength.
The \gls{fis} matrix $\boldsymbol{\Theta}$ is limited to be diagonal as it is a practical constraint commonly adopted in the \gls{ris} literature \cite{wu19,wu21}.
In addition, $\boldsymbol{\Theta}$ is assumed to be lossless and independent of the frequency, which are appropriate assumptions in this context that will be justified in the following subsection.

\subsection{Received Signal Power}

In this \gls{siso} system aided by a \gls{fis}, our goal is to maximize the received signal power
\begin{equation}
P_R=P_T\left\vert\mathbf{h}_{R}\boldsymbol{\Theta}\mathbf{h}_{T}-\mathbf{h}_{R}\mathbf{h}_{T}\right\vert^2,\label{eq:pr1-nlos}
\end{equation}
by fixing the \gls{fis} reflection coefficients in $\boldsymbol{\Theta}$ offline (as they are non-reconfigurable) and optimizing the wavelength $\lambda$ on a per-channel realization basis.
In detail, the wavelength $\lambda$ is optimized by solving
\begin{equation}
\underset{\lambda}{\mathsf{\mathrm{max}}}\;\;
P_T\left\vert\mathbf{h}_{R}\boldsymbol{\Theta}\mathbf{h}_{T}-\mathbf{h}_{R}\mathbf{h}_{T}\right\vert^2\;\;
\mathsf{\mathrm{s.t.}}\;\;
\eqref{eq:hR},\;\eqref{eq:hT},\;\boldsymbol{\Theta}\text{ is fixed},
\end{equation}
where $\boldsymbol{\Theta}$ is optimized offline and then fixed.
In the following, we show that this maximization problem can be globally solved in closed form by first deriving an upper bound on the received signal power $P_R$, and then proposing solutions for $\boldsymbol{\Theta}$ and $\lambda$ that achieve that upper bound.

The received signal power is upper bounded by
\begin{align}
P_R
&\leq P_T\left(\left\vert\mathbf{h}_{R}\boldsymbol{\Theta}\mathbf{h}_{T}\right\vert+\left\vert\mathbf{h}_{R}\mathbf{h}_{T}\right\vert\right)^2\label{eq:ineq1}\\
&\leq P_T\left(\left\Vert\mathbf{h}_{R}\right\Vert\left\Vert\boldsymbol{\Theta}\right\Vert\left\Vert\mathbf{h}_{T}\right\Vert+\left\Vert\mathbf{h}_{R}\right\Vert\left\Vert\mathbf{h}_{T}\right\Vert\right)^2\label{eq:ineq2}\\
&=4P_T\left\Vert\mathbf{h}_{R}\right\Vert^2\left\Vert\mathbf{h}_{T}\right\Vert^2=4P_TN^2,\label{eq:eq1}
\end{align}
where \eqref{eq:ineq1} follows from the triangle inequality, \eqref{eq:ineq2} applies the sub-multiplicity property of the $\ell_2$-norm and the Cauchy–Schwarz inequality, and \eqref{eq:eq1} holds since the spectral norm of any unitary matrix is one, i.e., $\Vert\boldsymbol{\Theta}\Vert=1$.
To achieve the upper bound $P_R^\star=4P_TN^2$, we fix $\boldsymbol{\Theta}=-\mathbf{I}$, giving a channel
$h=-2\mathbf{h}_{R}\mathbf{h}_{T}$,
and a received signal power
\begin{equation}
P_R=4P_T\left\vert\mathbf{h}_{R}\mathbf{h}_{T}\right\vert^2,\label{eq:pr2-nlos}
\end{equation}
and show in the following that $\lambda$ can be optimized on a per-channel realization basis to achieve $P_R^\star$.
Note that a \gls{fis} with $\boldsymbol{\Theta}=-\mathbf{I}$ can be interpreted as a \gls{ris} in which all elements are terminated with short-circuits instead of tunable loads, resulting in all phase shifts being fixed to $e^{j\theta_n}=-1$, for $n=1,\ldots,N$.
This choice ensures that the reflected path $\mathbf{h}_{R}\boldsymbol{\Theta}\mathbf{h}_{T}$ is always co-phased with the specular reflection path $-\mathbf{h}_{R}\mathbf{h}_{T}$, and hence they interfere constructively.
The optimality of $\boldsymbol{\Theta}=-\mathbf{I}$ is proved in the following, where we show that $P_R^\star$ can be achieved by optimizing $\lambda$.\footnote{While conventional \glspl{ris} typically exhibit lossy and frequency-dependent reflection matrices due to the use of tunable impedance loads such as varactors \cite{li21}, the proposed \gls{fis} architecture employs antenna elements that are short-circuited to ground.
Consequently, each element exhibits a reflection coefficient equal to $-1$ regardless of the frequency, which makes the assumptions of a lossless and frequency-independent reflection response reasonable in this context.}

By substituting \eqref{eq:hR} and \eqref{eq:hT} into \eqref{eq:pr2-nlos}, the received signal power is now
\begin{multline}
P_R
=4P_T\Bigg|\sum_{n=1}^N\left(e^{-j\frac{2\pi}{\lambda}\left[d_{R}-\left(n-\frac{N+1}{2}\right)d_A\sin(\theta_R)\right]}\right.\\
\left.\times e^{-j\frac{2\pi}{\lambda}\left[d_{T}-\left(n-\frac{N+1}{2}\right)d_A\sin(\theta_T)\right]}\right)\Bigg|^2.\label{eq:pr3-nlos}
\end{multline}
Thus, we achieve the upper bound $P_R^\star=4P_TN^2$ if and only if all the $n$ paths in \eqref{eq:pr3-nlos} are co-phased, i.e., they have the same phase up to a $2\pi$ factor.
This condition is formalized as
\begin{multline}
-\frac{2\pi}{\lambda}\left[d_{R}-\left(n-\frac{N+1}{2}\right)d_A\sin(\theta_R)\right.\\
\left.+d_{T}-\left(n-\frac{N+1}{2}\right)d_A\sin(\theta_T)\right]=\alpha+K_n2\pi,\label{eq:cond1-nlos}
\end{multline}
for $n=1,\ldots,N$, where $\alpha\in[0,2\pi)$ is an arbitrary phase and $K_n\in\mathbb{Z}$ is an integer varying with $n$.
By taking
\begin{equation}
\alpha=-\frac{2\pi}{\lambda}\left(d_{R}+d_{T}+\frac{N+1}{2}d_A\left(\sin\left(\theta_R\right)+\sin\left(\theta_T\right)\right)\right),
\end{equation}
condition \eqref{eq:cond1-nlos} simplifies as
\begin{equation}
\frac{n}{\lambda}d_A\left(\sin(\theta_R)+\sin(\theta_T)\right)=K_n,\label{eq:cond2-nlos}
\end{equation}
for $n=1,\ldots,N$.
Note that \eqref{eq:cond2-nlos} is satisfied for any $\lambda$ when $\sin(\theta_R)+\sin(\theta_T)=0$, i.e., $\theta_R=-\theta_T$, since the specular reflection (occurring for any frequency) is optimal in this case.
For $\theta_R\neq-\theta_T$, condition \eqref{eq:cond2-nlos} is satisfied by setting the wavelength as
$\lambda^\star(\theta_R,\theta_T)=nd_A(\sin(\theta_R)+\sin(\theta_T))/K_n$,
indicating that there are infinitely many possible values of $\lambda^\star$, depending on the values of $K_n$.
By selecting the largest among these values, since it corresponds to the lowest frequency $f^\star=c/\lambda^\star$, we obtain
$\lambda^\star(\theta_R,\theta_T)=d_A\vert\sin(\theta_R)+\sin(\theta_T)\vert$,
having a corresponding frequency
\begin{equation}
f^\star\left(\theta_R,\theta_T\right)=\frac{f_A}{\left\vert\sin(\theta_R)+\sin(\theta_T)\right\vert},\label{eq:f-nlos}
\end{equation}
where we have introduced $f_A=c/d_A$.

\subsection{Coverage}

\begin{figure}[t]
\centering
\includegraphics[width=0.32\textwidth]{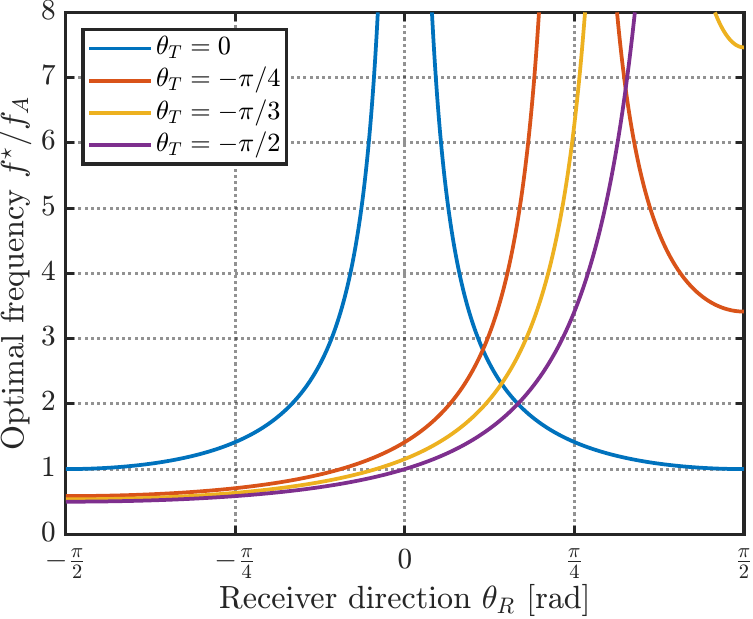}
\caption{Optimal frequency $f^\star$ normalized by $f_A=c/d_A$.}
\label{fig:f-nlos}
\end{figure}

\begin{figure*}[t]
\centering
\subfigure[$-\pi/2\leq\theta_T<\arcsin((1-W)/(1+W))$]{
\includegraphics[width=0.32\textwidth]{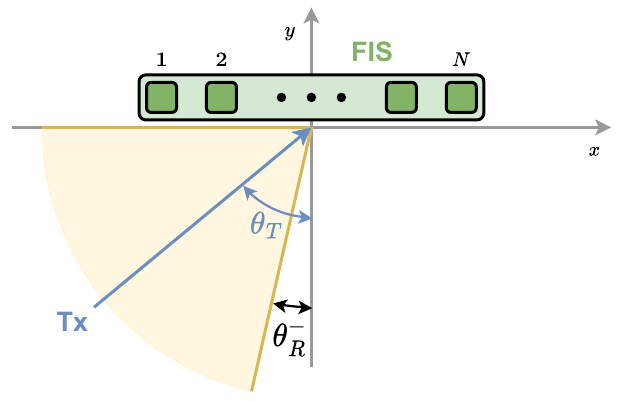}
\label{fig:system-nlos-coverage-a}}
\subfigure[$-\pi/2\leq\theta_T<\arcsin((1-W)/(1+W))$]{
\includegraphics[width=0.42\textwidth]{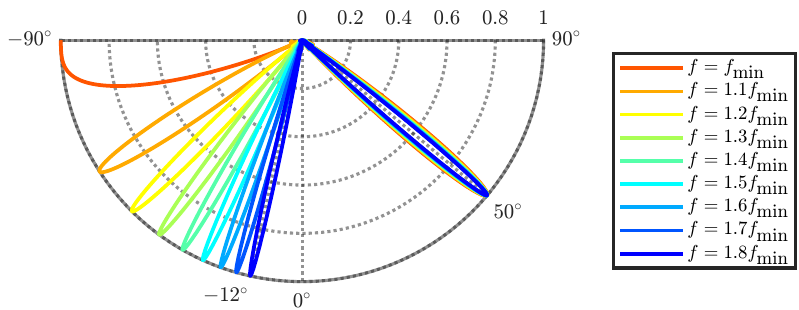}
\label{fig:system-nlos-coverage-b}}
\subfigure[$\arcsin((1-W)/(1+W))\leq\theta_T\leq0$]{
\includegraphics[width=0.32\textwidth]{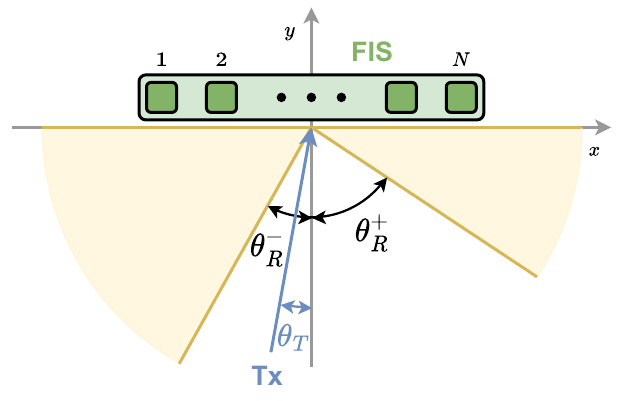}
\label{fig:system-nlos-coverage-c}}
\subfigure[$\arcsin((1-W)/(1+W))\leq\theta_T\leq0$]{
\includegraphics[width=0.42\textwidth]{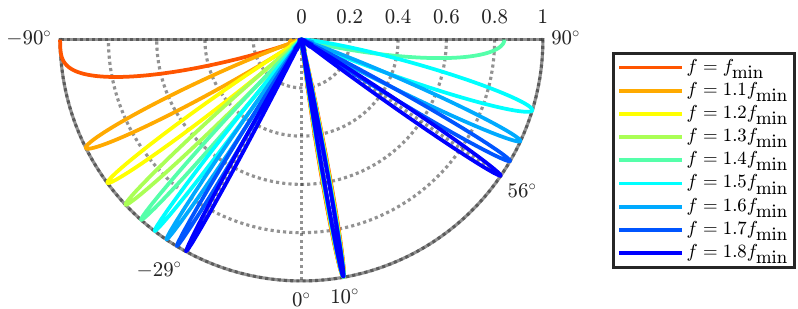}
\label{fig:system-nlos-coverage-d}}
\caption{FIS-aided system using movable signals with frequency range width $W=1.8$.
In (a)-(b), $\theta_T=-50^\circ$ gives $\theta_R^-=-12^\circ$ according to \eqref{eq:R-2-}.
(a) The region within coverage is highlighted in yellow.
(b) The radiation pattern is shown for different frequencies $f$.
In (c)-(d), $\theta_T=-10^\circ$ gives $\theta_R^-=-29^\circ$ and $\theta_R^+=56^\circ$ according to \eqref{eq:R-2-} and \eqref{eq:R-2+}.
(c) The region within coverage is highlighted in yellow.
(d) The radiation pattern is shown for different frequencies $f$.}
\end{figure*}

The optimal frequency $f^\star$ in \eqref{eq:f-nlos} could take any positive real value depending on $\theta_R$ and $\theta_T$, as shown in Fig.~\ref{fig:f-nlos}.
However, in practice, $f^\star$ cannot be reconfigured to any arbitrary value since only a limited frequency range is available.
As discussed in Section~\ref{sec:fundamentals-los} for the \gls{los} case, we consider the practical constraint of selecting $f^\star$ within an interval $[f_{\textrm{min}},f_{\textrm{max}}]$, where $f_{\textrm{max}}=Wf_{\textrm{min}}$ and $W>1$.
Thus, we now characterize the coverage of a frequency-domain \gls{sre} enabled by movable signals and \gls{fis}, defined as the portion of space where it is possible to achieve the upper bound $P_R^\star=4P_TN^2$ with a frequency constrained in $[f_{\textrm{min}},f_{\textrm{max}}]$.
To this end, we assume that the transmitter is a \gls{bs} at a fixed location, i.e., the angle $\theta_T$ is fixed.
Our goal is therefore to characterize the coverage as a function of $\theta_T$ and the frequency range width $W$.

Since the coverage is a symmetric function of $\theta_T$ given the symmetry of the system model, we focus on the case $\theta_T\leq0$.
To maximize the coverage, we set $f_{\textrm{min}}$ as the lowest $f^\star$ given by \eqref{eq:f-nlos}.
When $\theta_T\leq0$, \eqref{eq:f-nlos} is minimized in $\theta_R=-\pi/2$, yielding
\begin{equation}
f_{\textrm{min}}=f^\star\left(-\pi/2,\theta_T\right)=\frac{f_A}{1-\sin(\theta_T)},\label{eq:f-min-nlos}
\end{equation}
since $\left\vert\sin(\theta_R)+\sin(\theta_T)\right\vert=1-\sin(\theta_T)$ when $\theta_T\leq0$ and $\theta_R=-\pi/2$.
Note that the optimal frequency $f^\star$ in \eqref{eq:f-nlos} is a continuous function of $\theta_R$ which increases in the interval $[-\pi/2,-\theta_T)$.
Besides, $f^\star=f_{\textrm{min}}$ in $\theta_R=-\pi/2$ and $f^\star\to\infty$ as $\theta_R\to-\theta_T$.
Thus, there exists a value of $\theta_R$ in the interval $[-\pi/2,-\theta_T)$, denoted as $\theta_R^-$, for which $f^\star=f_{\textrm{max}}$, i.e.,
\begin{equation}
f^\star\left(\theta_R^-,\theta_T\right)=-\frac{f_A}{\sin(\theta_R^-)+\sin(\theta_T)}=f_{\textrm{max}},\label{eq:f-max-}
\end{equation}
following \eqref{eq:f-nlos}, since $\vert\sin(\theta_R^-)+\sin(\theta_T)\vert=-\sin(\theta_R^-)-\sin(\theta_T)$ for $\theta_T\leq0$ and $\theta_R^-\in[-\pi/2,-\theta_T)$.
By inverting \eqref{eq:f-max-}, $\theta_R^-$ is obtained as
\begin{align}
\theta_R^-
&=-\arcsin\left(\frac{f_A}{f_{\textrm{max}}}+\sin(\theta_T)\right)\label{eq:R-1-}\\
&=-\arcsin\left(\frac{1+(W-1)\sin(\theta_T)}{W}\right),\label{eq:R-2-}
\end{align}
where \eqref{eq:R-2-} is obtained by substituting $f_{\textrm{max}}=Wf_{\textrm{min}}$ in \eqref{eq:R-1-}, with $f_{\textrm{min}}$ given by \eqref{eq:f-min-nlos}.
Thus, for any $\theta_R\in[-\pi/2,\theta_R^-]$ we have $f^\star\in[f_{\textrm{min}},f_{\textrm{max}}]$ and the received signal power upper bound $P_R^\star=4P_TN^2$ can be achieved.

While $f^\star$ in \eqref{eq:f-nlos} is a continuous function of $\theta_R$ increasing in the interval $[-\pi/2,-\theta_T)$, it decreases in $(-\theta_T,\pi/2]$.
Thus, if $f^\star$ in $\theta_R=\pi/2$ is $f^\star\leq f_{\textrm{max}}$, i.e.,
\begin{equation}
f^\star\left(\pi/2,\theta_T\right)=\frac{f_A}{1+\sin(\theta_T)}\leq f_{\textrm{max}},\label{eq:cond-cov}
\end{equation}
there exists a value of $\theta_R$ in the interval $(-\theta_T,\pi/2]$, denoted as $\theta_R^+$, for which $f^\star=f_{\textrm{max}}$, i.e.,
\begin{equation}
f^\star\left(\theta_R^+,\theta_T\right)=\frac{f_A}{\sin(\theta_R^+)+\sin(\theta_T)}=f_{\textrm{max}}.\label{eq:f-max+}
\end{equation}
By inverting \eqref{eq:f-max+}, such a value $\theta_R^+$ is obtained as
\begin{align}
\theta_R^+
&=\arcsin\left(\frac{f_A}{f_{\textrm{max}}}-\sin(\theta_T)\right)\label{eq:R-1+}\\
&=\arcsin\left(\frac{1-(W+1)\sin(\theta_T)}{W}\right),\label{eq:R-2+}
\end{align}
where \eqref{eq:R-2+} is obtained by substituting $f_{\textrm{max}}=Wf_{\textrm{min}}$ in \eqref{eq:R-1+}, with $f_{\textrm{min}}$ given by \eqref{eq:f-min-nlos}.
In other words, if
\begin{equation}
\theta_T\geq\arcsin\left(\frac{1-W}{1+W}\right),
\end{equation}
which is a condition equivalent to \eqref{eq:cond-cov} since $f_{\textrm{max}}=Wf_{\textrm{min}}$ and $f_{\textrm{min}}$ is given by \eqref{eq:f-min-nlos}, for any $\theta_R\in[\theta_R^+,\pi/2]$ we have $f^\star\in[f_{\textrm{min}},f_{\textrm{max}}]$.

To summarize, when $\theta_T<\arcsin((1-W)/(1+W))$, the upper bound $P_R^\star=4P_TN^2$ can be achieved only for $\theta_R\in[-\pi/2,\theta_R^-]$, hence the fraction of space under coverage is $C=\theta_R^-+\pi/2$, as graphically shown in Fig.~\ref{fig:system-nlos-coverage-a} for the case $W=1.8$ and $\theta_T=-50^\circ$.
Within this region, it is always possible to create a grating lobe toward the intended direction depending on the frequency, as shown by the radiation pattern in Fig.~\ref{fig:system-nlos-coverage-b}, computed as $R(\theta_R)=\vert\mathbf{a}(\theta_R)\mathbf{a}(\theta_T)^T\vert^2$, where $\mathbf{a}(\theta)=[1,e^{j\frac{2\pi}{\lambda}d_A\sin(\theta)},\ldots,e^{j\frac{2\pi}{\lambda}d_A(N-1)\sin(\theta)}]/\sqrt{N}$, with $N=16$.
We also notice that the main beam is always generated toward $\theta_R=-\theta_T=50^\circ$, due to the specular reflection of the \gls{fis}.
Otherwise, when $\arcsin((1-W)/(1+W))\leq\theta_T\leq0$, the upper bound $P_R^\star=4P_TN^2$ can be achieved for $\theta_R\in[-\pi/2,\theta_R^-]\cup[\theta_R^+,\pi/2]$, and the coverage becomes $C=\pi+\theta_R^--\theta_R^+$, as graphically shown in Fig.~\ref{fig:system-nlos-coverage-c} for the case $W=1.8$ and $\theta_T=-10^\circ$.
Within this region, it is always possible to create a grating lobe toward the intended direction depending on the frequency, as shown by the radiation pattern in Fig.~\ref{fig:system-nlos-coverage-d}, while the main beam is always generated toward $\theta_R=-\theta_T=10^\circ$ due to the specular reflection.

\begin{figure}[t]
\centering
\includegraphics[width=0.32\textwidth]{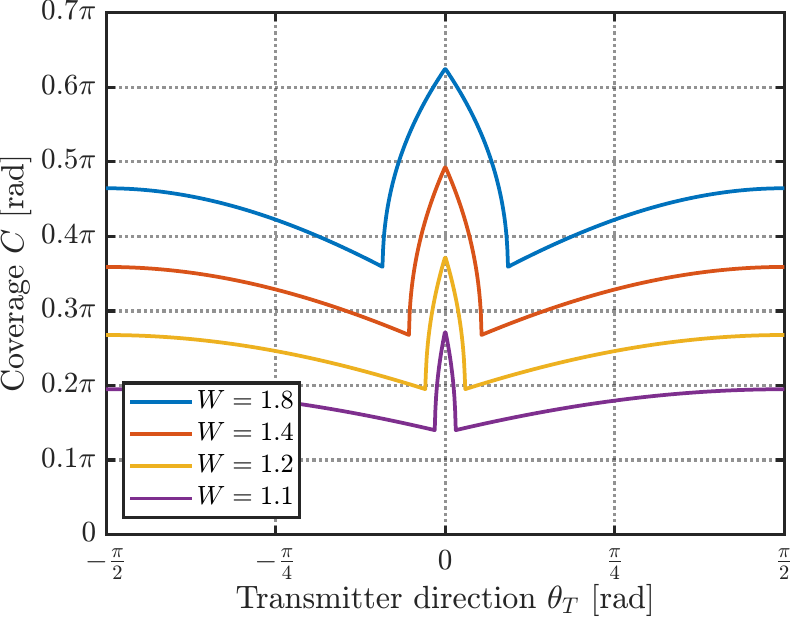}
\caption{Coverage $C$ versus the transmitter direction $\theta_T$ for different values of the frequency range width $W$.}
\label{fig:coverage-nlos}
\end{figure}

The coverage is therefore given by
\begin{equation}
C=
\begin{cases}
\theta_R^-+\pi/2 &,\;-\frac{\pi}{2}\leq\theta_T<\arcsin\left(\frac{1-W}{1+W}\right)\\
\pi+\theta_R^--\theta_R^+ &,\;\arcsin\left(\frac{1-W}{1+W}\right)\leq\theta_T\leq0
\end{cases},
\end{equation}
where $\theta_R^-$ and $\theta_R^+$ are given by \eqref{eq:R-2-} and \eqref{eq:R-2+} as functions of $W$ and $\theta_T$.
The coverage for $\theta_T>0$ can be readily obtained by exploiting the symmetry of $C$.
We report the coverage as a function of $W$ and $\theta_T$ in Fig.~\ref{fig:coverage-nlos}, which shows that the coverage increases with $W$, as observed for frequency-domain \glspl{sre} in \gls{los}.
In addition, we have $C<\pi$ in practice since $C=\pi$ can be achieved only with an infinite frequency range available.

\subsection{Comparison with Reconfigurable Intelligent Surfaces}
\label{sec:ris}

We have shown that the received signal power upper bound $P_R^\star=4P_TN^2$ can be achieved by deploying a \gls{fis} with phase shift matrix fixed to $\boldsymbol{\Theta}=-\mathbf{I}$, and optimizing the frequency of the transmitted signal as in \eqref{eq:f-nlos} depending on $\theta_R$ and $\theta_T$.
Note that a \gls{fis} is a surface composed of uniformly spaced scattering elements, and thus differs from a simple metallic plate.
Compared to a \gls{ris}, this solution exhibits lower losses since it eliminates the need for tunable impedance components such as PIN diodes and varactors, and has significantly reduced hardware complexity and cost, as it is non-reconfigurable.

Since we have shown that a \gls{fis}-aided system with movable signals can achieve a received signal power $P_R^\star=4P_TN^2$, we compare this value with the maximum received signal power achievable in a \gls{ris}-aided system where $\boldsymbol{\Theta}$ is reconfigurable and the signal frequency is fixed.
With a \gls{ris}, the received signal power maximization problem writes as
\begin{align}
\underset{\boldsymbol{\Theta}}{\mathsf{\mathrm{max}}}\;\;
&P_T\left\vert\mathbf{h}_{R}\boldsymbol{\Theta}\mathbf{h}_{T}-\mathbf{h}_{R}\mathbf{h}_{T}\right\vert^2\label{eq:prob-obj-ris1}\\
\mathsf{\mathrm{s.t.}}\;\;\;
&\mathbf{h}_{R},\mathbf{h}_{T}\text{ are fixed},\;
\boldsymbol{\Theta}=\text{diag}\left(e^{j\theta_1},\ldots,e^{j\theta_N}\right),\label{eq:prob-c2-ris1}
\end{align}
indicating that the phase shift matrix $\boldsymbol{\Theta}$ is optimized given the fixed channel realizations $\mathbf{h}_{R}$ and $\mathbf{h}_{T}$.
It is well-known from previous work \cite{wu19,wu21} that problem \eqref{eq:prob-obj-ris1}-\eqref{eq:prob-c2-ris1} can be globally solved by setting
\begin{gather}
\boldsymbol{\Theta}^\star=\text{diag}\left(e^{j\theta_1^\star},\ldots,e^{j\theta_N^\star}\right),\label{eq:Theta-star}\\
\theta_n^\star=-\arg\left(\left[\mathbf{h}_R\right]_n\left[\mathbf{h}_T\right]_n\right)+\arg\left(-\mathbf{h}_R\mathbf{h}_T\right),\;\forall n,
\end{gather}
giving the maximum received signal power as
\begin{align}
P_R^\star
&=P_T\left(\sum_{n=1}^N\left\vert\left[\mathbf{h}_{R}\right]_n\left[\mathbf{h}_{T}\right]_n\right\vert+\left\vert\mathbf{h}_{R}\mathbf{h}_{T}\right\vert\right)^2\\
&=P_T\left(N+\left\vert\mathbf{h}_{R}\mathbf{h}_{T}\right\vert\right)^2,\label{eq:pr-ris}
\end{align}
where \eqref{eq:pr-ris} holds since $\vert[\mathbf{h}_{R}]_n[\mathbf{h}_{T}]_n\vert=1$, for $n=1,\ldots,N$.\footnote{The upper bound derived in \eqref{eq:ineq1}-\eqref{eq:eq1} is valid also in a \gls{ris}-aided system.
We have indeed $P_T(N+\vert\mathbf{h}_{R}\mathbf{h}_{T}\vert)^2\leq4P_TN^2$ since $\vert\mathbf{h}_{R}\mathbf{h}_{T}\vert\leq N$ following the Cauchy–Schwarz inequality.}

By taking the expectation of \eqref{eq:pr-ris} assuming random $\theta_R$ and $\theta_T$ independent and uniformly distributed as $\theta_R,\theta_T,\sim\mathcal{U}[-\pi/2,\pi/2]$, we have
\begin{equation}
\mathbb{E}\left[P_R^\star\right]
=P_T\left(N^2+2N\mathbb{E}\left[\left\vert\mathbf{h}_{R}\mathbf{h}_{T}\right\vert\right]+\mathbb{E}\left[\left\vert\mathbf{h}_{R}\mathbf{h}_{T}\right\vert^2\right]\right),\label{eq:gain-ris-mean1}
\end{equation}
and by considering the approximation $\mathbf{h}_{R}\mathbf{h}_{T}\sim\mathcal{CN}(0,N)$ (which gives $\mathbb{E}[\vert\mathbf{h}_{R}\mathbf{h}_{T}\vert]=\sqrt{\frac{\pi}{4}N}$ and $\mathbb{E}[\vert\mathbf{h}_{R}\mathbf{h}_{T}\vert^2]=N$), we obtain
\begin{equation}
\mathbb{E}\left[P_R^\star\right]
=P_T\left(N^2+\sqrt{\pi N}N+N\right),
\end{equation}
which scales with $N^2$, in agreement with \cite{wu19,wu21}.

\begin{figure}[t]
\centering
\includegraphics[width=0.32\textwidth]{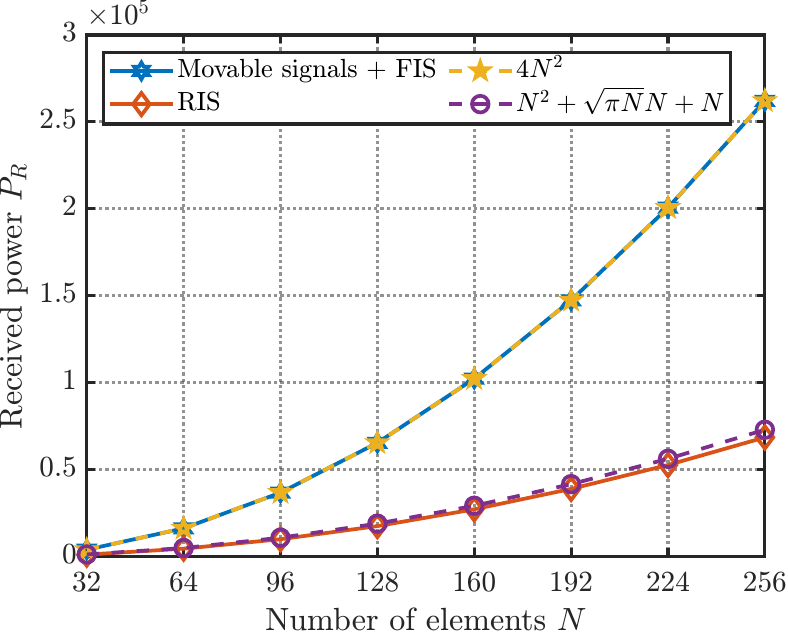}
\caption{Average received signal power obtained with FIS and RIS.}
\label{fig:pr-scaling}
\end{figure}

In Fig.~\ref{fig:pr-scaling}, we report the received signal power obtained by a \gls{fis}-aided system with movable signals and a \gls{ris}-aided system.
In the \gls{fis}-aided system, the \gls{fis} phase shift matrix is fixed to $\boldsymbol{\Theta}=-\mathbf{I}$ and the signal frequency is set to $f^\star$ as in \eqref{eq:f-nlos}.
In the \gls{ris}-aided system, the \gls{ris} phase shift matrix is set to $\boldsymbol{\Theta}^\star$ as in \eqref{eq:Theta-star}, i.e., it is optimized assuming ideal and arbitrary \gls{ris} reconfiguration, and the signal frequency is fixed to $f=c/(2d_A)$, such that $d_A$ is half-wavelength.
The simulated average received signal power and the corresponding theoretical scaling laws are reported, where $P_T=1$ and $d_A$, $d_R$, and $d_T$ can have arbitrary values in the simulations as they do not impact the results.
We observe that the theoretical scaling law for \gls{fis} is exact and the one for \gls{ris} is approximately accurate.
Remarkably, in the regime $N\to\infty$, \gls{fis} provides an average received signal power four times higher than \gls{ris}, which only gives $\mathbb{E}[P_R^\star]\approx P_TN^2$.
The reason for this is that in a \gls{fis}-aided system with movable signals, by reconfiguring the wavelength, we control the channels $\mathbf{h}_{R}$ and $\mathbf{h}_{T}$ and hence both terms $\mathbf{h}_{R}\boldsymbol{\Theta}\mathbf{h}_{T}$ and $-\mathbf{h}_{R}\mathbf{h}_{T}$ in the channel expression \eqref{eq:h1-nlos}.
Conversely, in a \gls{ris}-aided system, we reconfigure the \gls{ris} phase shift matrix $\boldsymbol{\Theta}$, and therefore we can only control the term $\mathbf{h}_{R}\boldsymbol{\Theta}\mathbf{h}_{T}$ and have no impact on $-\mathbf{h}_{R}\mathbf{h}_{T}$.
Since a \gls{fis} is non-reconfigurable, it also offers enormous benefits over \gls{ris} in terms of hardware complexity and cost, control overhead, and ohmic losses.

\section{Performance Evaluation of\\Frequency-Domain SREs}
\label{sec:performance}

In Section~\ref{sec:fundamentals-los}, we have introduced how frequency-domain \glspl{sre} can improve \gls{los} communications through movable signals.
In Section~\ref{sec:fundamentals-nlos}, we have investigated the benefits of frequency-domain \glspl{sre} in \gls{nlos}, showing the benefits of movable signals used with a \gls{fis} that reflects them from the transmitter to the receiver.
To gain fundamental insights, we have considered \gls{los} fading channels neglecting the path gain, we have assumed perfect \gls{csi}, and we have reconfigured the frequency with full flexibility.
However, the path gain decreases with frequency, which negatively impacts the received signal power when higher frequencies are used.
Acquiring perfect \gls{csi} is challenging due to the non-reconfigurability of the precoder $\mathbf{w}$ and the \gls{fis} phase shift matrix $\boldsymbol{\Theta}$ in the two considered scenarios, respectively.
Furthermore, the operating frequency can only take a finite number of values within a frequency range of limited width in practical systems, due to spectrum allocation policies.
In this section, we address these issues by proposing a practical transmission protocol for movable signals that applies to arbitrary channel realizations.
Our protocol selects the optimal operating frequency from a limited set of possible values without requiring explicit \gls{csi}.
Under these more practical assumptions, the performance of movable signals is compared with \gls{egt} in the \gls{los} scenario, and with \gls{ris} in the \gls{nlos} scenario.

\subsection{Transmission Protocol}
\label{sec:protocol}

Consider a single-user system where movable signals are employed.
For example, this can be the \gls{miso} system in Fig.~\ref{fig:system-los}, where the precoder of the multi-antenna transmitter is fixed to $\mathbf{w}=\mathbf{1}/\sqrt{N}$, or the \gls{siso} system in Fig.~\ref{fig:system-nlos}, where the \gls{fis} matrix is fixed to $\boldsymbol{\Theta}=-\mathbf{I}$.
To optimize the signal frequency without explicitly knowing the channel, the following protocol can be adopted.
We assume that the available frequency range $[f_{\textrm{min}},f_{\textrm{max}}]$ is uniformly divided into $S$ subchannels, each spaced by bandwidth $B=(f_{\textrm{max}}-f_{\textrm{min}})/(S-1)$.
Thus, we denote as $f_s=f_{\textrm{min}}+(s-1)B$ the center frequency of the $s$th subchannel, for $s=1,\ldots,S$, such that the first and last frequencies are $f_1=f_{\textrm{min}}$ and $f_S=f_{\textrm{max}}$, respectively.
The goal is to identify the center frequency that maximizes the received signal power, and perform the communication in the corresponding subchannel.
To this end, we propose a three-stage transmission protocol as follows.
\begin{enumerate}
\item The transmitter sends a pilot signal in each of the $S$ subchannels.
\item The receiver assesses the received power of the $S$ pilot signals and identifies the index $s^\star$ of the strongest one.
Then, it feeds back the index $s^\star$ to the transmitter.
\item The transmitter performs the actual transmission operating in the $s^\star$th subchannel, i.e., using the frequency $f_{s^\star}$ as the center frequency.
\end{enumerate}
Note that such a protocol applies to any channel realization, in both \gls{los} and \gls{nlos} conditions, and it enables the selection of the optimal frequency without explicitly acquiring the \gls{csi}.
The pilot overhead in terms of frequency resources scales linearly with the number of subchannels, i.e., $\mathcal{O}(S)$.
All pilot signals across the $S$ subchannels can be transmitted simultaneously, so the associated overhead is confined to a single signaling interval and remains extremely limited in time.
Since the optimal frequency is selected by comparing the received signal power over all possible $S$ subchannels, the computational complexity of the proposed protocol is $\mathcal{O}(S)$, growing linearly with $S$.

\begin{figure}[t]
\centering
\includegraphics[width=0.32\textwidth]{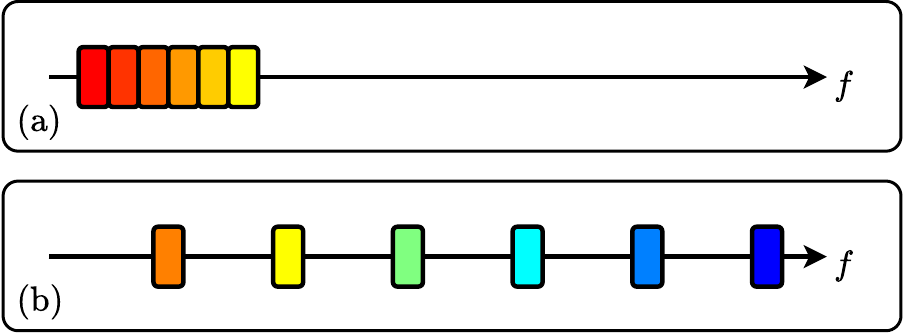}
\caption{Spectrum allocated to an operator (a) in current networks and (b) with movable signals.}
\label{fig:spectrum}
\end{figure}

Movable signals require the frequency to be reconfigured over a wide range, and therefore a different spectrum allocation to operators.
In current 4G and 5G networks, an operator is typically allocated a few large, contiguous spectrum blocks, which are then subdivided into resource units to serve multiple users.
For example, the spectrum allocation in the UK is available at \cite{ofc26}.
This frequency allocation is qualitatively represented in Fig.~\ref{fig:spectrum}(a), showing that the spectrum available to an operator is a large contiguous block divided into six resource units.
To enable movable signals, the available spectrum must be more fragmented and distributed over a wide frequency range, as shown in Fig.~\ref{fig:spectrum}(b), where the six available resource units are better distributed over the frequency axis.
In this way, the operator can select the operating frequency for a specific user with the sufficient flexibility required by movable signals.
Observe that the two spectrum allocations in Fig.~\ref{fig:spectrum}(a) and Fig.~\ref{fig:spectrum}(b) have the same total bandwidth.
Therefore, each operator can maintain the same total bandwidth, only differently allocated.
In addition, interference with other operators (and in general other services) is avoided, since operators are still allocated orthogonal resources.

\begin{figure*}[t]
\centering
\subfigure[$W=1.1$]{\includegraphics[width=0.24\textwidth]{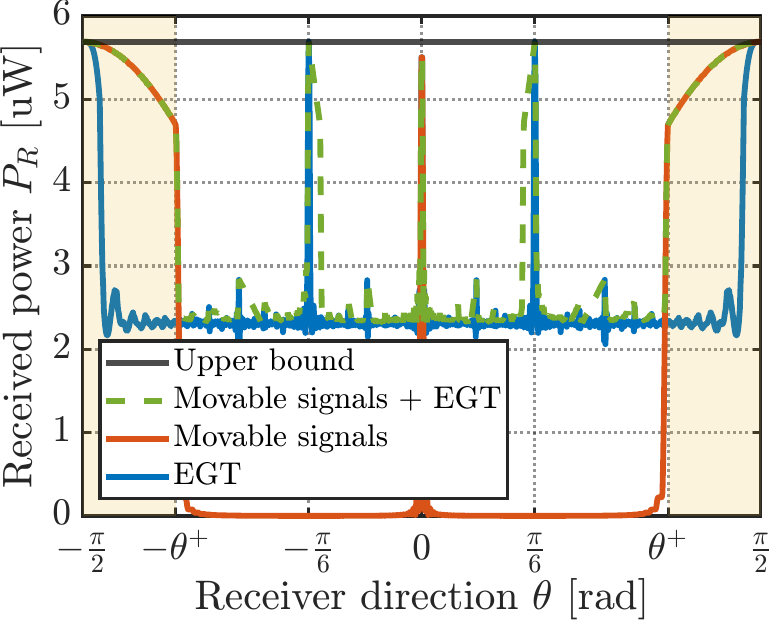}}
\subfigure[$W=1.2$]{\includegraphics[width=0.24\textwidth]{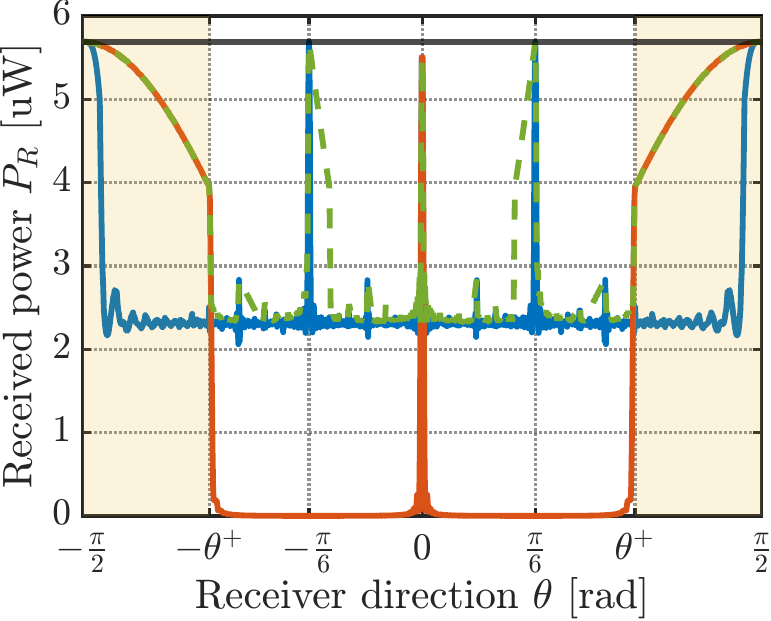}}
\subfigure[$W=1.4$]{\includegraphics[width=0.24\textwidth]{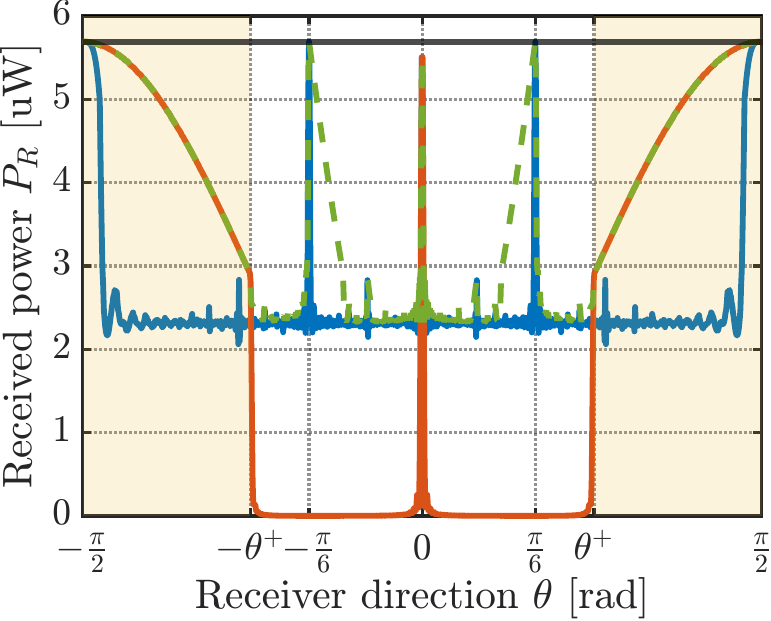}}
\subfigure[$W=1.8$]{\includegraphics[width=0.24\textwidth]{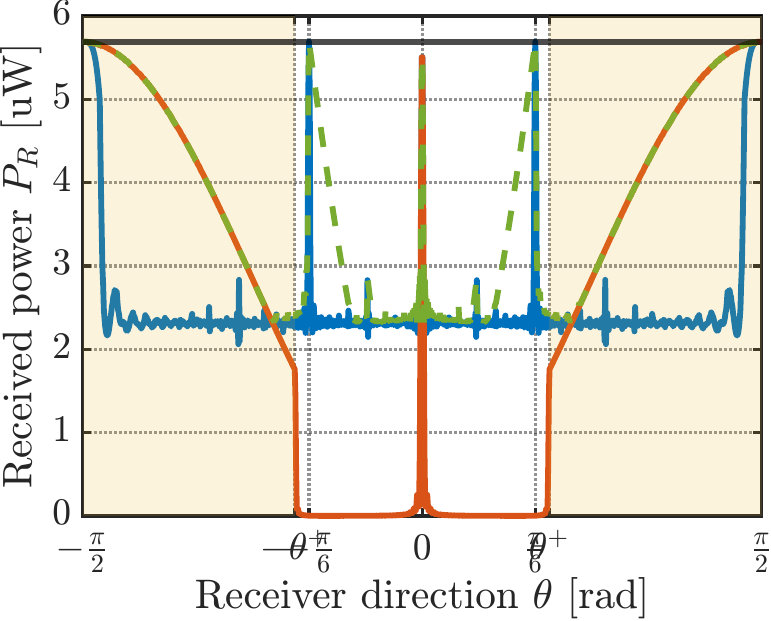}}
\caption{Received power versus the receiver direction $\theta$ for different values of the frequency range width $W$.}
\label{fig:pr-los}
\end{figure*}

\begin{figure}[t]
\centering
\includegraphics[width=0.32\textwidth]{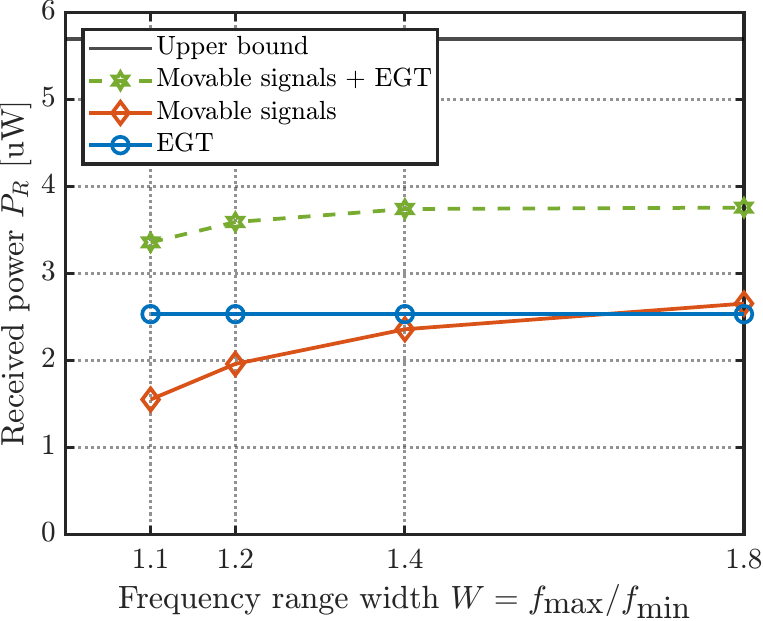}
\caption{Average received power versus the frequency range width $W$.}
\label{fig:pr-w-los}
\end{figure}

\subsection{Frequency-Domain SREs in LoS}

Given the proposed transmission protocol applied to the \gls{miso} system in Fig.~\ref{fig:system-los}, we numerically evaluate the performance of movable signals and compare it with the performance of \gls{egt}.
To account for the impact of the frequency on the path gain, the channel between the transmitter and receiver is modeled as $\mathbf{h}=\sqrt{\rho}\tilde{\mathbf{h}}$, where $\rho$ is the path gain and $\tilde{\mathbf{h}}$ is the small-scale fading effects.
The path gain $\rho$ is given by $\rho=(4\pi d/\lambda)^{-2}$ depending on the distance $d$ and the wavelength $\lambda$,
and the small-scale fading $\tilde{\mathbf{h}}$ is given by \eqref{eq:h-los}.

We consider a numerical setup with $N=64$ transmitting antennas, $P_T=1$~W, $d=10$~m, and $f_{\textrm{min}}=8$~GHz, resulting in an antenna spacing $d_A=c/f_A=c/f_{\textrm{min}}$ because of \eqref{eq:f-min-los}.
In Fig.~\ref{fig:pr-los}, we report the received signal power versus the angle $\theta\in[-\pi/2,\pi/2]$ achieved by \gls{egt} and movable signals for different values of the frequency interval width $W\in\{1.1,1.2,1.4,1.8\}$, where the number of subchannels is set as $S\in\{128,256,512,1024\}$, respectively.
We consider the following four baselines.

\textit{First}, ``EGT'' gives the performance of \gls{egt}, assuming that it is realized with \gls{rf} phase shifters that are lossless and quantized with one-bit resolution, i.e., can have a phase $0$ or $\pi$.
The signal frequency is fixed to $f=f_{\textrm{min}}$, which is the value that maximizes the path gain $\rho$.
Formally, the precoder $\mathbf{w}$ is optimized by solving
\begin{equation}
\underset{\mathbf{w}}{\mathsf{\mathrm{max}}}\;\;
P_T\left\vert\mathbf{h}\mathbf{w}\right\vert^2\;\;
\mathsf{\mathrm{s.t.}}\;\;
\mathbf{h}\text{ is fixed},\;\left[\mathbf{w}\right]_n\in\left\{\pm\frac{1}{\sqrt{N}}\right\},\;\forall n.\label{eq:prob-egt2}
\end{equation}
Since globally solving problem \eqref{eq:prob-egt2} requires an exhaustive search over $2^{N}$ possible precoders, we solve it through a closed-form solution that is numerically shown to approximately perform as the global optimal. 
Noticing that the received signal power is maximized when all the $N$ additive terms in the scalar product $\mathbf{h}\mathbf{w}=\sum_{n=1}^N[\mathbf{w}]_n[\mathbf{h}]_n$ in \eqref{eq:prob-egt2} are co-phased, we ensure that they all have positive real part by setting
$[\mathbf{w}]_n=\sign(\Re([\mathbf{h}]_n))/\sqrt{N}$,
for $n=1,\ldots,N$.
In other words, this solution gives $[\mathbf{w}]_n=1/\sqrt{N}$ or $-1/\sqrt{N}$ if $[\mathbf{h}]_n$ has positive or negative real part, respectively, such that the $N$ additive terms in $\mathbf{h}\mathbf{w}=\sum_{n=1}^N[\mathbf{w}]_n[\mathbf{h}]_n$ have all positive real part.

\textit{Second}, ``Movable signals'' gives the performance of a frequency-domain \gls{sre} with movable signals, where the precoder is fixed to $\mathbf{w}=\mathbf{1}/\sqrt{N}$ and the frequency is optimized through the protocol described in Section~\ref{sec:protocol}.

\textit{Third}, ``Movable signals + EGT'' gives the performance obtained by optimizing both the signal frequency $f$ and the precoder $\mathbf{w}$.
For each center frequency, the precoder $\mathbf{w}$ is optimized as $[\mathbf{w}]_n=\sign(\Re([\mathbf{h}]_n))/\sqrt{N}$, and the center frequency giving the highest received signal power is selected for transmission.\footnote{In the baselines ``EGT'' and ``Movable signals + EGT'', we assume perfect channel knowledge to optimize the precoder, while in ``Movable signals'' the adopted protocol does not require perfect knowledge of the channel.}

\textit{Fourth}, ``Upper bound'' gives the upper bound on the received signal power achievable in the case of maximum path gain $\rho$, which is $P_R\leq\rho_{\textrm{max}}P_TN$, where $\rho_{\textrm{max}}$ is the maximum path gain, i.e., computed with $\lambda=c/f_{\textrm{min}}$.
This upper bound does not vary with the receiver direction $\theta$.

From Fig.~\ref{fig:pr-los}, we make four observations.
\textit{First}, \gls{egt} achieves a received signal power approximately constant across $\theta$, with five spikes in $\theta\in\{-\pi/2,-\pi/6,0,\pi/6,\pi/2\}$ since in these angles we can achieve maximum performance with a precoder having phases constrained to be $0$ or $\pi$.
\textit{Second}, movable signals achieve very strong received signal power for $\theta$ within the coverage, i.e., for $\theta\in[-\pi/2,-\theta^+]\cup[\theta^+,\pi/2]$, and for $\theta=0$, while very weak received signal power elsewhere.
\textit{Third}, the region of space within coverage increases with the frequency range width $W$.
\textit{Fourth}, by jointly employing movable signals and \gls{egt}, i.e., by jointly optimizing the frequency and the precoder, we can achieve a received signal power greater than the maximum achievable by using either movable signals or \gls{egt} alone.

In Fig.~\ref{fig:pr-w-los}, we report the received signal power averaged over the angle $\theta\in[-\pi/2,\pi/2]$ achieved by the four considered baselines versus the frequency range width $W$.
We observe that the performance of movable signals increases with $W$, since a larger $W$ provides a wider frequency range for selecting the optimal frequency.
Interestingly, when $W=1.8$, movable signals achieve higher performance than \gls{egt}.

\begin{figure*}[t]
\centering
\subfigure[$\theta_T=0$, $W=1.1$]{\includegraphics[width=0.24\textwidth]{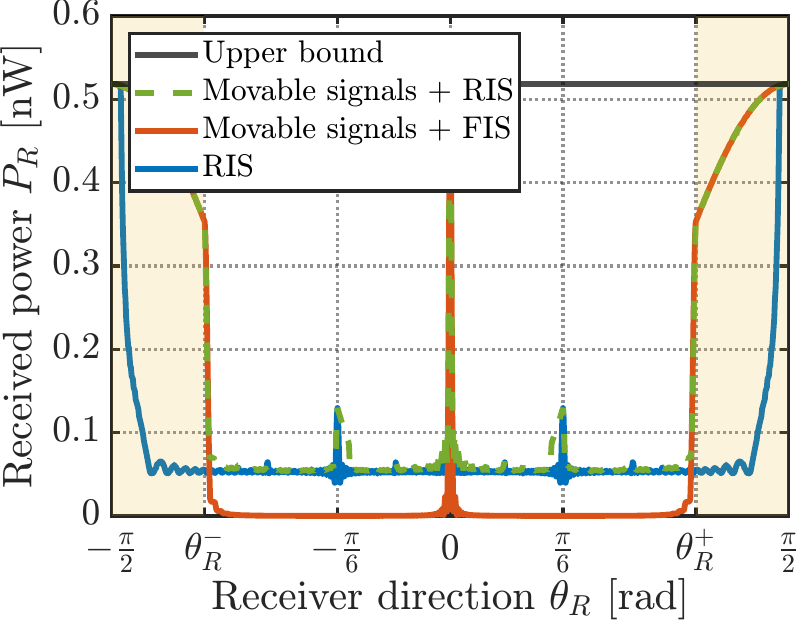}}
\subfigure[$\theta_T=0$, $W=1.2$]{\includegraphics[width=0.24\textwidth]{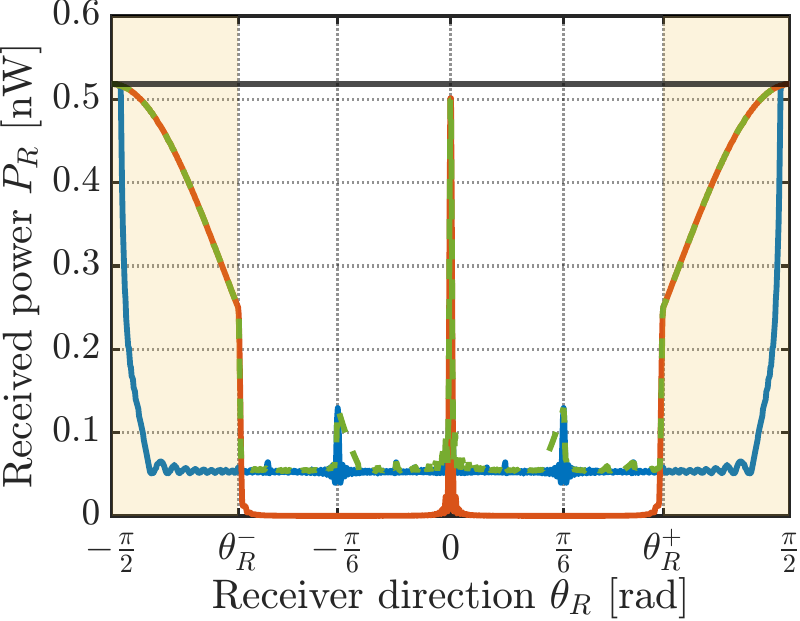}}
\subfigure[$\theta_T=0$, $W=1.4$]{\includegraphics[width=0.24\textwidth]{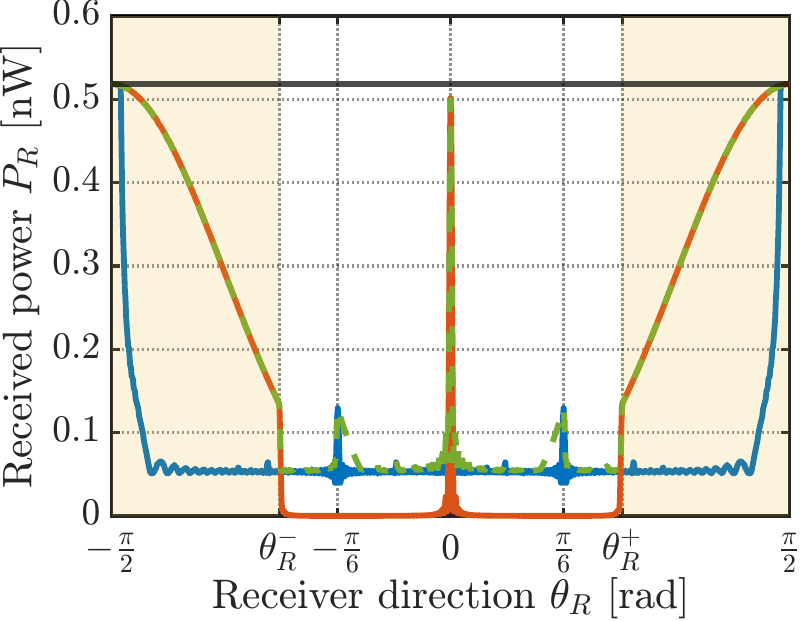}}
\subfigure[$\theta_T=0$, $W=1.8$]{\includegraphics[width=0.24\textwidth]{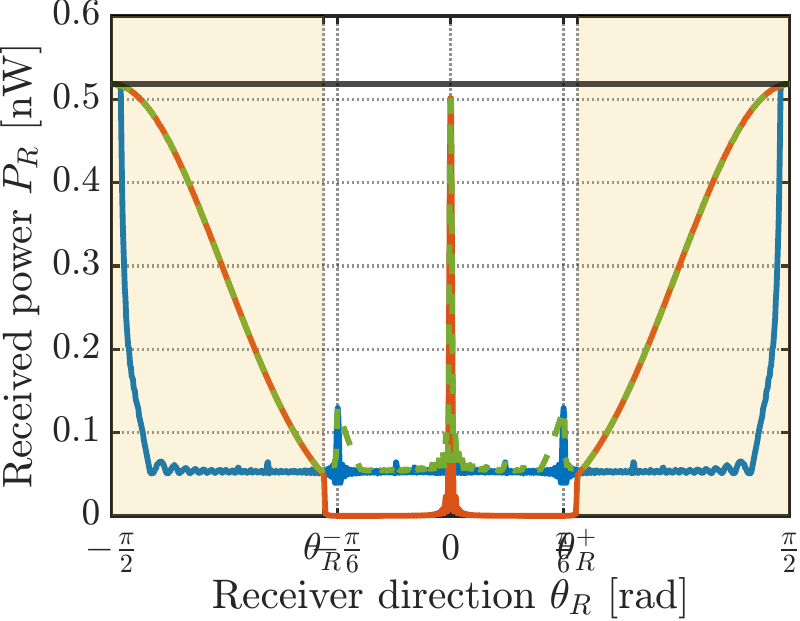}}
\subfigure[$\theta_T=-\pi/2$, $W=1.1$]{\includegraphics[width=0.24\textwidth]{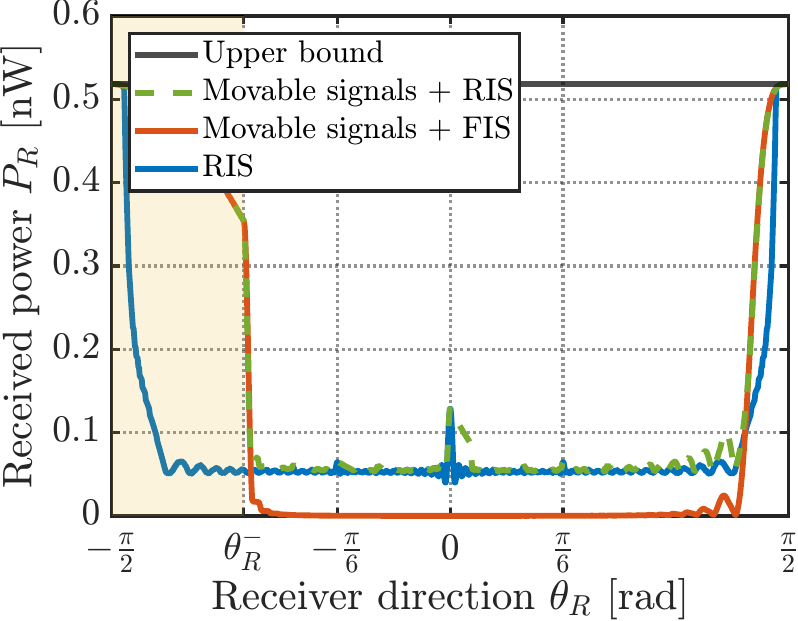}}
\subfigure[$\theta_T=-\pi/2$, $W=1.2$]{\includegraphics[width=0.24\textwidth]{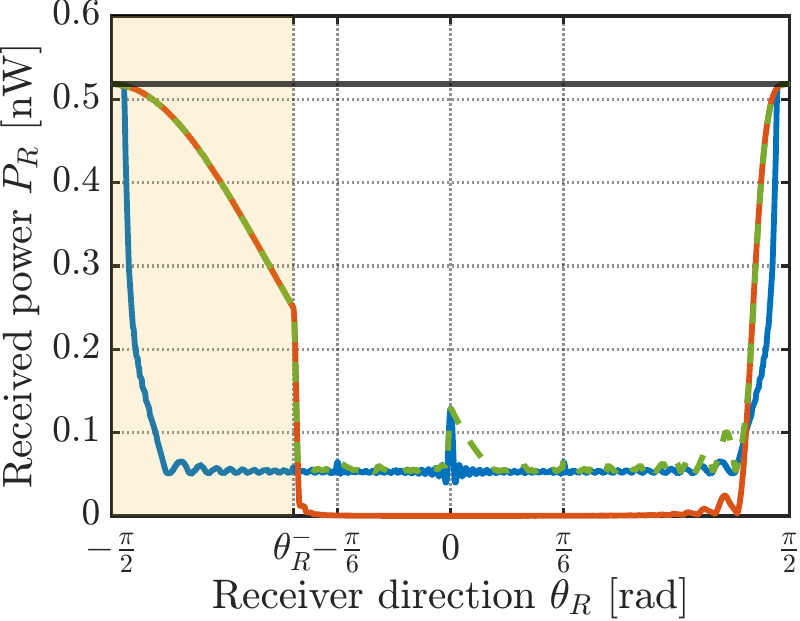}}
\subfigure[$\theta_T=-\pi/2$, $W=1.4$]{\includegraphics[width=0.24\textwidth]{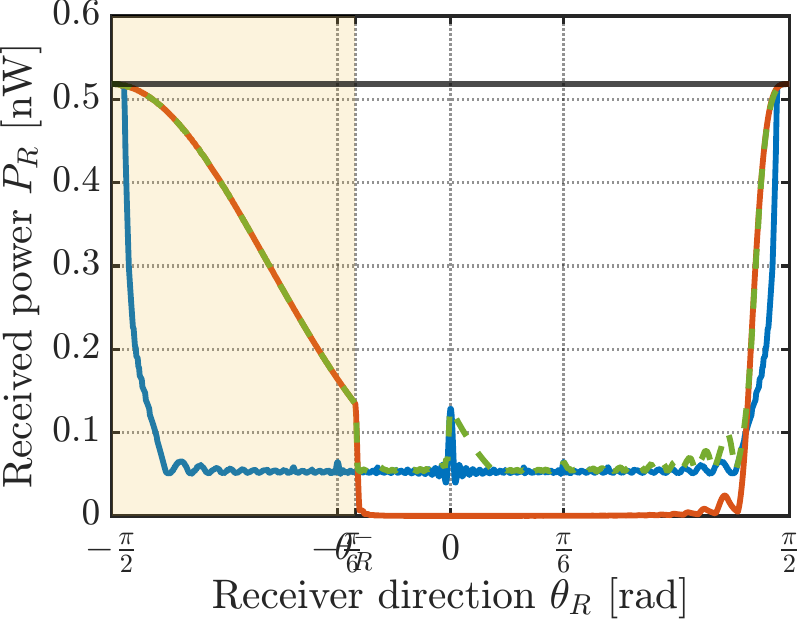}}
\subfigure[$\theta_T=-\pi/2$, $W=1.8$]{\includegraphics[width=0.24\textwidth]{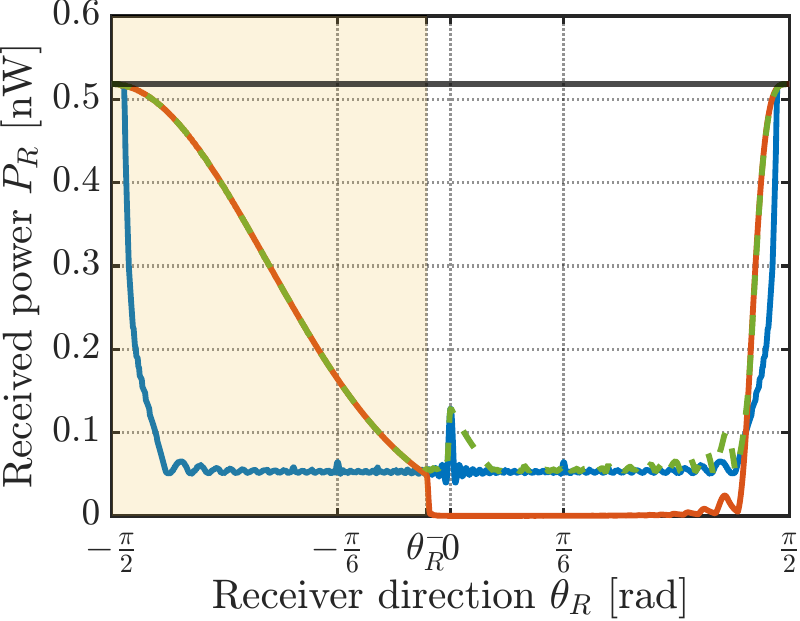}}
\caption{Received power versus the receiver direction $\theta_R$ for different values of the transmitter direction $\theta_T$ and frequency range width $W$.}
\label{fig:pr-nlos}
\end{figure*}

\begin{figure}[t]
\centering
\subfigure[$\theta_T=0$]{\includegraphics[width=0.32\textwidth]{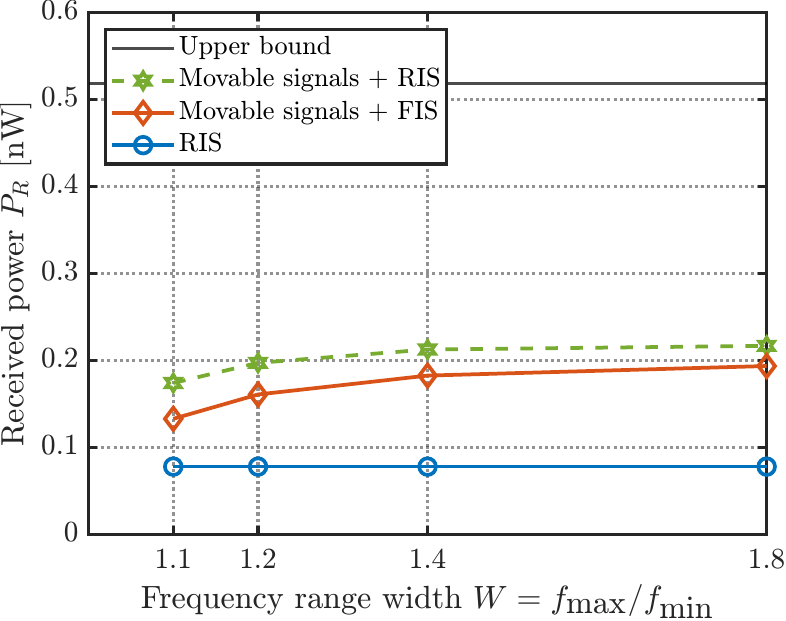}}
\subfigure[$\theta_T=-\pi/2$]{\includegraphics[width=0.32\textwidth]{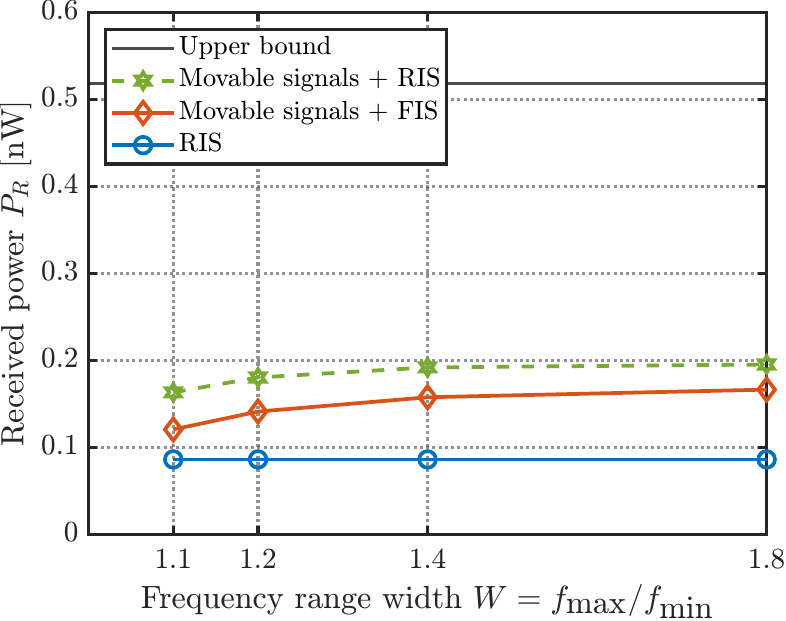}}
\caption{Average received power versus the frequency range width $W$.}
\label{fig:pr-w-nlos}
\end{figure}

\subsection{Frequency-Domain SREs in NLoS}

Considering the \gls{nlos} \gls{siso} system in Fig.~\ref{fig:system-nlos}, we numerically evaluate the performance of movable signals with \gls{fis} and compare it with the performance of \gls{ris}.
The channel between the transmitter and receiver is given by $h=-2\mathbf{h}_{R}\mathbf{h}_{T}$ (since $\boldsymbol{\Theta}=-\mathbf{I}$), where the channels $\mathbf{h}_R$ and $\mathbf{h}_T$ are modeled as $\mathbf{h}_R=\sqrt{\rho_{R}}\tilde{\mathbf{h}}_R$ and $\mathbf{h}_T=\sqrt{\rho_{T}}\tilde{\mathbf{h}}_T$, where $\rho_X$ are the path gains and $\tilde{\mathbf{h}}_X$ are the small-scale fading effects, for $X\in\{R,T\}$.
The path gains $\rho_{R}$ and $\rho_{T}$ are given as a function of the distances $d_R$ and $d_T$, respectively, and the wavelength $\lambda$ as
$\rho_X=(4\pi d_X/\lambda)^{-2}$,
for $X\in\{R,T\}$,
and the small-scale fading $\tilde{\mathbf{h}}_{R}$ and $\tilde{\mathbf{h}}_{T}$ are modeled as in \eqref{eq:hR} and \eqref{eq:hT}, respectively.

We consider a setup with the following numerology.
The transmitter is assumed to be a fixed \gls{bs}, whose direction is fixed, while the receiver direction is uniformly distributed, i.e., $\theta_R\in[-\pi/2,\pi/2]$.
We set $N=64$ elements, $P_T=1$~W, $d_R=5$~m, $d_T=10$~m, and $f_{\textrm{min}}=8$~GHz, resulting in an antenna spacing $d_A=c/f_A=c/(f_{\textrm{min}}(1-\sin(\theta_T)))$ following \eqref{eq:f-min-nlos}.
In Fig.~\ref{fig:pr-nlos}, we report the received signal power versus the receiver direction $\theta_R\in[-\pi/2,\pi/2]$ achieved by \gls{ris} and movable signals with \gls{fis} for different values of the transmitter direction $\theta_T\in\{0,-\pi/2\}$ and frequency interval width $W\in\{1.1,1.2,1.4,1.8\}$, where the number of subchannels is respectively set as $S\in\{128,256,512,1024\}$.
The following four baselines are compared.

\textit{First}, ``RIS'' shows the performance of a \gls{ris}-aided system, where the \gls{ris} is assumed to be lossless, and each \gls{ris} element can be controlled with one bit of resolution, i.e., $\theta_n\in\left\{0,\pi\right\}$, for $n=1,\ldots,N$, for practical reasons.
The frequency of the signal is fixed to $f=f_{\textrm{min}}$.
In detail, the \gls{ris} phase shift matrix is optimized as
\begin{align}
\underset{\boldsymbol{\Theta}}{\mathsf{\mathrm{max}}}\;\;
&P_T\left\vert\mathbf{h}_{R}\boldsymbol{\Theta}\mathbf{h}_{T}-\mathbf{h}_{R}\mathbf{h}_{T}\right\vert^2\label{eq:prob-obj-ris2}\\
\mathsf{\mathrm{s.t.}}\;\;\;
&\mathbf{h}_{R},\mathbf{h}_{T}\text{ are fixed},\\
&\boldsymbol{\Theta}=\text{diag}\left(e^{j\theta_1},\ldots,e^{j\theta_N}\right),\;\theta_n\in\left\{0,\pi\right\}.\label{eq:prob-c2-ris2}
\end{align}
Problem \eqref{eq:prob-obj-ris2}-\eqref{eq:prob-c2-ris2} can be solved with an empirical closed-form solution with strong performance.
Note that the received signal power is maximized when all the $N$ additive terms in the product $\mathbf{h}_{R}\boldsymbol{\Theta}\mathbf{h}_{T}=\sum_{n=1}^N[\mathbf{h}_{R}]_n[\boldsymbol{\Theta}]_{n,n}[\mathbf{h}_{T}]_n$ in \eqref{eq:prob-obj-ris2} are co-phased with $-\mathbf{h}_{R}\mathbf{h}_{T}$.
Thus, we can ensure that their real part all have the same sign as $\Re(-\mathbf{h}_{R}\mathbf{h}_{T})$ by setting
\begin{equation}
\left[\boldsymbol{\Theta}\right]_{n,n}=
\sign\left(\Re\left(\left[\mathbf{h}_R\right]_n\left[\mathbf{h}_T\right]_n\right)\right)
\sign\left(\Re\left(-\mathbf{h}_R\mathbf{h}_T\right)\right),\label{eq:ris-opt}
\end{equation}
for $n=1,\ldots,N$.
With this solution, the $N$ additive terms in $\mathbf{h}_{R}\boldsymbol{\Theta}\mathbf{h}_{T}=\sum_{n=1}^N[\mathbf{h}_{R}]_n[\boldsymbol{\Theta}]_{n,n}[\mathbf{h}_{T}]_n$ have positive real part if $-\mathbf{h}_{R}\mathbf{h}_{T}$ has positive real part, or negative real part otherwise.

\textit{Second}, ``Movable signals + FIS'' shows the performance of movable signals used with a \gls{fis} with phase shift matrix fixed to $\boldsymbol{\Theta}=-\mathbf{I}$, where the signal frequency is optimized as described in Section~\ref{sec:protocol}.

\textit{Third}, ``Movable signals + RIS'' indicates the performance achieved by movable signals used with \gls{ris}. We jointly optimize the signal frequency $f$ and the \gls{ris} phase shift matrix $\boldsymbol{\Theta}$ by computing the optimal $\boldsymbol{\Theta}$ as in \eqref{eq:ris-opt} for any available frequency, and the best frequency with its optimal $\boldsymbol{\Theta}$ is selected for transmission.\footnote{In the baselines ``RIS'' and ``Movable signals + RIS'', the \gls{ris} is optimized assuming perfect channel knowledge, while in ``Movable signals'' we do not require perfect channel knowledge.}

\textit{Fourth}, ``Upper bound'' is the upper bound on the received signal power given by $P_R\leq4\rho_{R,\textrm{max}}\rho_{T,\textrm{max}}P_TN^2$, where $\rho_{R,\textrm{max}}$ and $\rho_{T,\textrm{max}}$ are the path gains when $\lambda=c/f_{\textrm{min}}$.

From Fig.~\ref{fig:pr-nlos}, we make four observations.
\textit{First}, \gls{ris} achieves a performance approximately constant across $\theta_R$, and reaches the performance upper bound in $\theta_R\in\{-\pi/2,\pi/2\}$ since in these angles the optimal $\boldsymbol{\Theta}^\star$ has phases constrained to $0$ or $\pi$.
\textit{Second}, movable signals with \gls{fis} achieve very strong received signal power for $\theta_R$ within the coverage, i.e., for $\theta_R\in[-\pi/2,\theta_R^-]\cup[\theta_R^+,\pi/2]$ if $\theta_T=0$, and for $\theta_R\in[-\pi/2,\theta_R^-]$ if $\theta_T=-\pi/2$.
A very weak received signal power is achieved elsewhere.
\textit{Third}, the coverage region increases with the frequency range width $W$.
\textit{Fourth}, by using movable signals with \gls{ris}, we achieve a received signal power greater than the maximum achievable by using either aided movable signals with \gls{fis} or \gls{ris} with a fixed signal frequency.

In Fig.~\ref{fig:pr-w-nlos}, we report the received signal power averaged over the receiver direction $\theta_R\in[-\pi/2,\pi/2]$ versus the frequency range width $W$.
As expected, the performance of movable signals increases with $W$, both when they are used with \gls{fis} or with \gls{ris}, as a larger $W$ implies more flexibility in choosing the operating frequency.
Interestingly, movable signals with \gls{fis} achieve a much higher performance than \gls{ris} with fixed-frequency signals, even for a frequency range width as low as $W=1.1$.
This is because movable signals fully exploit the specular reflection of the surface, by aligning the two additive terms $\mathbf{h}_{R}\boldsymbol{\Theta}\mathbf{h}_{T}$ and $-\mathbf{h}_{R}\mathbf{h}_{T}$ in the channel expression \eqref{eq:h1-nlos}, as discussed in Section~\ref{sec:ris}.
These numerical results indicate the benefit of reconfiguring the signal frequency rather than the phase shift matrix $\boldsymbol{\Theta}$, leading also to a much simplified hardware infrastructure.

\section{Conclusion}
\label{sec:conclusion}

\Glspl{sre} have advanced wireless communications by allowing us to reconfigure the wireless channel.
While previous literature focused on \gls{em}-domain and space-domain \glspl{sre}, in this paper, we show that \glspl{sre} can also be enabled in the frequency domain.
To this end, we propose movable signals as a technique where the frequency of the signal can be optimized depending on the channel realization.
Through analytical and numerical analyses, we demonstrate that movable signals provide substantial gains in received signal power under both \gls{los} and \gls{nlos} conditions.
In \gls{los}, they can outperform quantized \gls{egt} when the signal frequency can be reconfigured with sufficient flexibility.
In \gls{nlos}, movable signals benefit from reflective surfaces with fixed \gls{em} properties deployed in the propagation environment, denoted as \gls{fis}.
Movable signals in \gls{fis}-aided systems can achieve up to a fourfold increase in received signal power compared to \gls{ris}-aided systems using fixed-frequency signals.
Furthermore, these gains are realized without electronically tunable or physically movable hardware.

Future research avenues include, but are not limited to, the following four topics.
\textit{First}, investigate how multi-user systems can benefit from frequency-domain \glspl{sre}.
To this end, algorithms to serve multiple users through movable signals need to be developed, as well as multiple access techniques.
In \cite{ner26-2}, it has been shown that movable signals can ``orthogonalize'' the channels in two-user systems, thereby radically suppressing interference and enabling large capacity regions.
\textit{Second}, design the fixed parameters of a \gls{sre} depending on the statistics of the propagation environment.
For instance, the precoder at the transmitter $\mathbf{w}$, the phase shift matrix of the \gls{fis} $\boldsymbol{\Theta}$, and the array geometries can be optimized offline to maximize the ergodic performance.
A first study in this direction has shown that in dual-polarized systems it is beneficial to design \glspl{fis} with non-diagonal reflection matrices \cite{ner26-1}.
The center frequencies of the available subchannels can also be optimized offline, and the use of carrier aggregation can be considered to extend the coverage.
\textit{Third}, explore multi-domain \glspl{sre} to enhance communications by jointly reconfiguring \gls{em}-properties, distances, and the frequency.
While previous work considered \glspl{sre} enabled jointly in the \gls{em} and space domains (see Fig.~\ref{fig:enablers}), the frequency domain can be added to the picture.
\textit{Fourth}, design and performance assessment of systems aided by movable signals and \gls{fis} in the presence of hardware non-idealities such as: \gls{nlos} multipath effects, antenna impedance mismatch, mutual coupling between antennas, imperfect frequency synchronization, non-white noise, and losses and frequency-dependent behaviors at the \gls{fis}.

\bibliographystyle{IEEEtran}
\bibliography{IEEEabrv,main}

\end{document}